%% file: AdaptiveDFSLAM-LiDeuLeiMey_arXiv_V2.tex
\documentclass[journal]{IEEEtran}
\usepackage[table]{xcolor}

\newcommand{\exportFigures}{false} 
\input{./definitions.tex}
\newcommand{\tikzfolder}{./compiledPlots/}
\input{pgfplot}

\IEEEoverridecommandlockouts

\begin{document}
	\allowdisplaybreaks
	\frenchspacing
	\title{\huge Adaptive Multipath-Based SLAM\\[.3mm] for Distributed MIMO Systems}
	\author{
	Xuhong~Li$^\ast$, \textit{Member,~IEEE}, 
	Benjamin~J.~B.~Deutschmann$^\ast$, \textit{Student~Member,~IEEE}, \\ 
	Erik~Leitinger, \textit{Member,~IEEE}, 
	Florian~Meyer, \textit{Member,~IEEE} \vspace*{-9mm}
	
	\thanks{Xuhong~Li is with the Department of Electrical and Information Technology, Lund University, Sweden (email: \texttt{xuhong.li@eit.lth.se}). This work was performed at University of California San Diego, CA, USA. Benjamin~J.~B.~Deutschmann and Erik~Leitinger are with the Institute of Communication Networks and Satellite Communications, Graz University of Technology, Austria (email: \texttt{$\{$benjamin.deutschmann, erik.leitinger$\}$@tugraz.at}). Florian~Meyer is with the Department of Electrical and Computer Engineering and the Scripps Institution of Oceanography, University of California San Diego, CA, USA (email: \texttt{flmeyer@ucsd.edu}).
	This work was supported by the Knut and Alice Wallenberg Foundation, by the Ericsson Research Foundation, by AMBIENT-6G project of the European Commission under Grant No. 101192113, and by the National Science Foundation (NSF) under CAREER Award No. 2146261.
	
	Xuhong~Li and Benjamin J. B. Deutschmann contributed equally as co-first authors to this work.}
	}

	\maketitle
	\renewcommand{\baselinestretch}{1} 

	\begin{abstract}
		\subfile{./InputFiles/abstract}
	\end{abstract}

	\IEEEpeerreviewmaketitle
	\acresetall
	\vspace*{-1mm}
	\section{Introduction}
	\label{sec:intro}
	\subfile{./InputFiles/Introduction}

	\section{Geometrical Model of the Environment}
	\label{sec:GeometricalRelations}

	\subfile{./InputFiles/GeometricalRelations}

	\section{System model}
	\label{sec:SystemModel}
	\subfile{./InputFiles/SystemModel}

	\section{Problem Formulation and Factor Graph}
	\label{sec:ProblemFormulation}
	\subfile{./InputFiles/ProblemFormulation}

	\section{Sum Product Algorithm}
	\label{sec:SPA}
	\subfile{./InputFiles/SPA}

	\section{Experimental Results}
	\label{sec:ExperimentalResults}
	\subfile{./InputFiles/ExperimentalResults}

	\section{Conclusions}
	\label{sec:Conclusions}
	\subfile{./InputFiles/Conclusions}

	\bibliographystyle{IEEEtran}
    \balance
	\bibliography{IEEEabrv,references}

	\includepdf[pages=-]{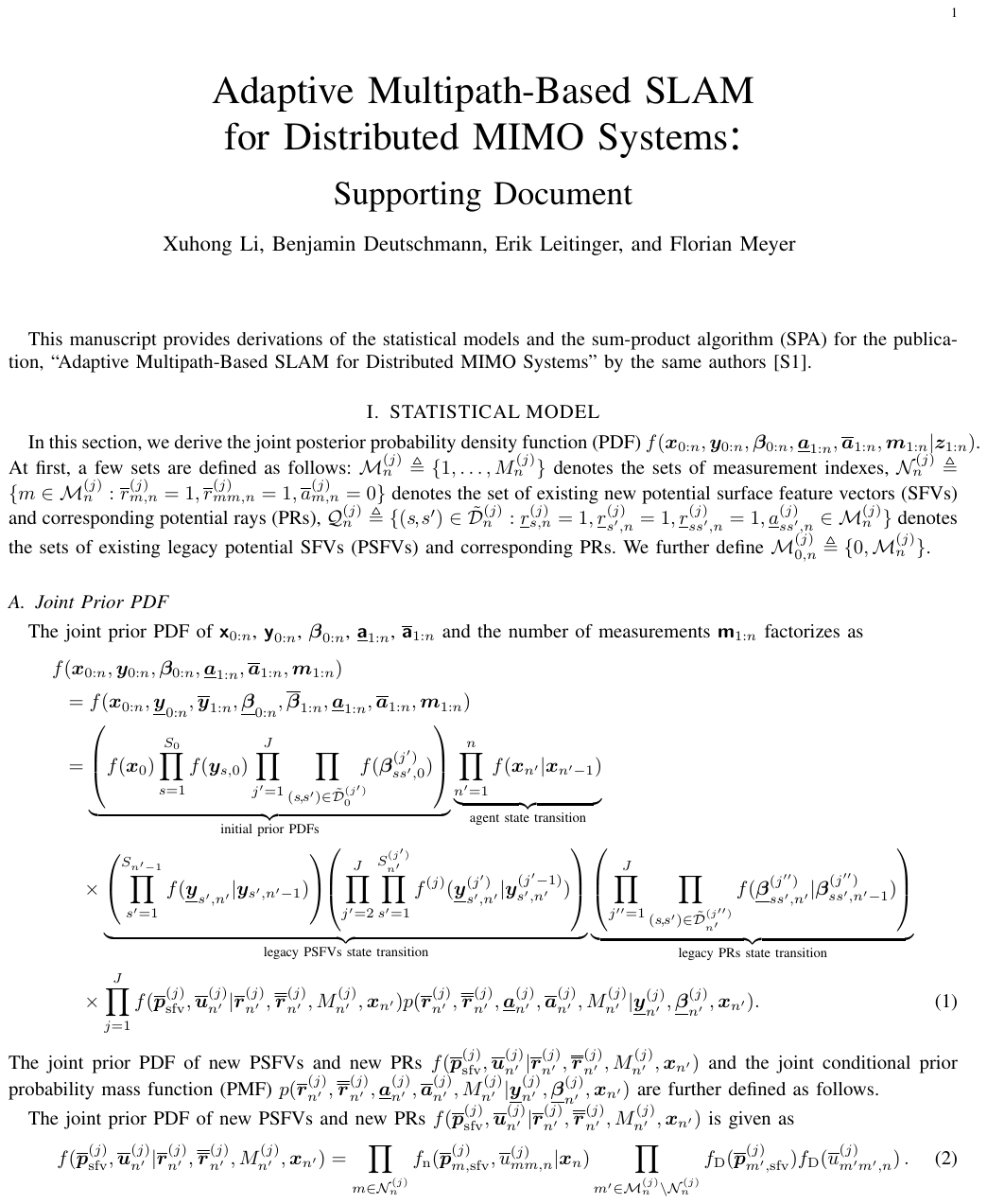}

\end{document}

%% file: definitions.tex
\usepackage{tikz}
\usetikzlibrary{backgrounds}
\tikzstyle{load}   = [ultra thick,-latex]
\tikzstyle{stress} = [-latex]
\tikzstyle{dim}    = [latex-latex]
\tikzstyle{axis}   = [-latex,black!55]

\definecolor{green(pigment)}{rgb}{0.0, 0.65, 0.31}
\definecolor{frenchblue}{rgb}{0.0, 0.45, 0.73} 
\definecolor{mediumcandyapplered}{rgb}{0.89, 0.02, 0.17}
\definecolor{darkGreyGreen598081}{rgb}{0.349, 0.502, 0.506}
\definecolor{darkGreyGreen2b6a99}{rgb}{0.349, 0.502, 0.506}
\definecolor{darkBlue1b7c3d}{rgb}{0.14, 0.35, 0.51}

\definecolor{redb11927}{rgb}{0.734, 0.104, 0.162}
\definecolor{blue4695be}{rgb}{0.171, 0.364, 0.465}

\usepackage{tkz-euclide}
\usetikzlibrary{arrows.meta}
\usepackage{setspace}
\usepackage{bm}
\usepackage{mathtools}
\usepackage{graphicx}
\usepackage{relsize}
\usepackage{cite}
\usepackage{amsmath}
\usepackage{color}
\usepackage{psfrag}
\usepackage{amssymb}
\usepackage{dblfloatfix}
\usepackage[font=small]{caption}
\usepackage{tabularx}
\usepackage{makecell}
\usepackage{lipsum}
\usepackage{float}
\usepackage{subfiles}
\usepackage{epstopdf}
\usepackage{verbatim}
\usepackage{amsmath}
\usepackage{ifthen}
\usepackage{pgfplots}
\usepackage{pgfplotstable}
\usepackage{tikzscale}
\usepackage{acronym}
\usepackage{amsfonts}
\usepackage{tikz-3dplot}
\usepackage{cuted}
\usepackage{subfig}
\usepackage{balance}          
\usepackage{pdfpages}  

\usetikzlibrary{intersections}   
\usepgfplotslibrary{fillbetween}

\usetikzlibrary{spy,backgrounds}
\usetikzlibrary{fit}
\usetikzlibrary{calc}

\usetikzlibrary{3d}
\usetikzlibrary{shapes}
\usetikzlibrary {patterns,patterns.meta}
\usepackage{pgffor}
\usetikzlibrary{shapes.geometric}
\usetikzlibrary{angles,quotes}

\renewcommand{\baselinestretch}{1} 

\colorlet{veccol}{green!50!black}
\colorlet{projcol}{blue!70!black}
\colorlet{myblue}{blue!80!black}
\colorlet{myred}{red!90!black}
\colorlet{mydarkblue}{blue!50!black}
\tikzset{>=latex} 
\tikzstyle{proj}=[projcol!80,line width=0.08] 
\tikzstyle{area}=[draw=veccol,fill=veccol!80,fill opacity=0.6]
\tikzstyle{vector}=[-stealth,myblue,thick,line cap=round]
\tikzstyle{unit vector}=[->,veccol,thick,line cap=round]
\tikzstyle{dark unit vector}=[unit vector,veccol!70!black]
\usetikzlibrary{angles,quotes} 

\definecolor{RDlightgreen}{RGB}{141 192 69}
\definecolor{RDgreen}{rgb}{0.3647, 0.4275, 0.2667}
\definecolor{RDdarkgreen}{rgb}{0.2196, 0.2196, 0.2196}
\definecolor{RDmaroon}{rgb}{.522,.22,.353} %

\definecolor{IEEEblue}{RGB}{0 98 155}
\definecolor{IEEElightblue}{RGB}{0 181 226}
\definecolor{IEEEturquoise}{RGB}{0 156 166}
\definecolor{IEEEred}{RGB}{186 12 47}
\definecolor{IEEEgreen}{RGB}{0 132 61}
\definecolor{IEEElightgreen}{RGB}{120 190 32}
\definecolor{IEEEorange}{RGB}{225 163 0}
\definecolor{IEEEyellow}{RGB}{255 209 0}
\definecolor{IEEEviolett}{RGB}{152 29 151}
\definecolor{IEEEdarkmaroon}{RGB}{134 31 65}
\definecolor{IEEEdarkorange}{RGB}{232 119 34}

\def\sfv{{sfv}}
\def\pr{{pr}}
\def\ip{{IP}}

\newcommand{\ist}{\hspace*{.3mm}}
\newcommand{\rmv}{\hspace*{-.3mm}}
\newcommand{\iist}{\hspace*{1mm}}
\newcommand{\rrmv}{\hspace*{-1mm}}
\newcommand{\nn}{\nonumber}

\providecommand{\norm}[1]{\lVert#1\rVert}

\DeclareMathAlphabet{\mathsfbr}{OT1}{cmss}{m}{n}
\SetMathAlphabet{\mathsfbr}{bold}{OT1}{cmss}{bx}{n}
\DeclareRobustCommand{\msf}[1]{%
	\ifcat\noexpand#1\relax\msfgreek{#1}\else\mathsfbr{#1}\fi
}

\makeatletter
\newcommand{\msfgreek}[1]{\csname s\expandafter\@gobble\string#1\endcsname}
\makeatother

\DeclareRobustCommand{\mcal}[1]{%
	\ifcat\noexpand#1\relax\mathnormal{#1}\else\cal{#1}\fi
}
\DeclareRobustCommand{\BM}[1]{%
	\ifcat\noexpand#1\relax\bm{\boldUppercaseItalicGreek{#1}}\else\bm{#1}\fi
}
\makeatletter
\newcommand{\boldUppercaseItalicGreek}[1]{\csname var\expandafter\@gobble\string#1\endcsname}
\makeatother

\newcommand{\rv}[1]{\msf{#1}}
\newcommand{\RV}[1]{\bm{\msf{#1}}}

\newcommand{\V}[1]{\bm{#1}}

\DeclareMathVersion{timesmath}
\SetSymbolFont{letters}{timesmath}{OML}{ntxmi}{m}{it}
\DeclareMathVersion{timesmathbold}
\SetSymbolFont{letters}{timesmathbold}{OML}{ntxmi}{b}{it}

\makeatletter
\newif\ifAC@uppercase@first%
\def\Aclp#1{\AC@uppercase@firsttrue\aclp{#1}\AC@uppercase@firstfalse}%
\def\AC@aclp#1{%
	\ifcsname fn@#1@PL\endcsname%
	\ifAC@uppercase@first%
	\expandafter\expandafter\expandafter\MakeUppercase\csname fn@#1@PL\endcsname%
	\else%
	\csname fn@#1@PL\endcsname%
	\fi%
	\else%
	\AC@acl{#1}s%
	\fi%
}%
\def\Acp#1{\AC@uppercase@firsttrue\acp{#1}\AC@uppercase@firstfalse}%
\def\AC@acp#1{%
	\ifcsname fn@#1@PL\endcsname%
	\ifAC@uppercase@first%
	\expandafter\expandafter\expandafter\MakeUppercase\csname fn@#1@PL\endcsname%
	\else%
	\csname fn@#1@PL\endcsname%
	\fi%
	\else%
	\AC@ac{#1}s%
	\fi%
}%
\def\Acfp#1{\AC@uppercase@firsttrue\acfp{#1}\AC@uppercase@firstfalse}%
\def\AC@acfp#1{%
	\ifcsname fn@#1@PL\endcsname%
	\ifAC@uppercase@first%
	\expandafter\expandafter\expandafter\MakeUppercase\csname fn@#1@PL\endcsname%
	\else%
	\csname fn@#1@PL\endcsname%
	\fi%
	\else%
	\AC@acf{#1}s%
	\fi%
}%
\def\Acsp#1{\AC@uppercase@firsttrue\acsp{#1}\AC@uppercase@firstfalse}%
\def\AC@acsp#1{%
	\ifcsname fn@#1@PL\endcsname%
	\ifAC@uppercase@first%
	\expandafter\expandafter\expandafter\MakeUppercase\csname fn@#1@PL\endcsname%
	\else%
	\csname fn@#1@PL\endcsname%
	\fi%
	\else%
	\AC@acs{#1}s%
	\fi%
}%
\edef\AC@uppercase@write{\string\ifAC@uppercase@first\string\expandafter\string\MakeUppercase\string\fi\space}%
\def\AC@acrodef#1[#2]#3{%
	\@bsphack%
	\protected@write\@auxout{}{%
		\string\newacro{#1}[#2]{\AC@uppercase@write #3}%
	}\@esphack%
}%
\def\Acl#1{\AC@uppercase@firsttrue\acl{#1}\AC@uppercase@firstfalse}
\def\Acf#1{\AC@uppercase@firsttrue\acf{#1}\AC@uppercase@firstfalse}
\def\Ac#1{\AC@uppercase@firsttrue\ac{#1}\AC@uppercase@firstfalse}
\def\Acs#1{\AC@uppercase@firsttrue\acs{#1}\AC@uppercase@firstfalse}
\robustify\Aclp
\robustify\Acfp
\robustify\Acp
\robustify\Acsp
\robustify\Acl
\robustify\Acf
\robustify\Ac
\robustify\Acs
\makeatother

\makeatletter
\def\underbracex#1#2{\mathop{\vtop{\m@th\ialign{##\crcr
				$\hfil\displaystyle{#2}\hfil$\crcr
				\noalign{\kern3\p@\nointerlineskip}%
				#1\crcr\noalign{\kern3\p@}}}}\limits}

\def\upbracefilla{$\m@th \setbox\z@\hbox{$\braceld$}%
	\bracelu\leaders\vrule \@height\ht\z@ \@depth\z@\hfill 
	\leaders\vrule \@height\ht\z@ \@depth\z@\hfill\bracerd
	\braceld\leaders\vrule \@height\ht\z@ \@depth\z@\hfill
	\kern\p@\vrule \@width\p@\kern\p@\vrule \@width\p@\kern\p@\vrule \@width\p@ 
	$}

\def\upbracefilld{$\m@th \setbox\z@\hbox{$\braceld$}%
	\vrule \@width\p@\kern\p@\vrule \@width\p@\kern\p@\vrule \@width\p@\kern\p@
	\leaders\vrule \@height\ht\z@ \@depth\z@\hfill\braceru$}

\makeatother

\acrodef{2d}[2D]{two-dimensional}
\acrodef{3d}[3D]{three-dimensional}
\acrodef{5g}[5G]{5th-generation}
\acrodef{pnt}[PNT]{positioning, navigation and timing}
\acrodef{pa}[PA]{physical anchor}
\acrodef{bs}[BS]{base station}
\acrodef{awgn}[AWGN]{additive white Gaussian noise}
\acrodef{va}[VA]{virtual anchor}
\acrodef{mpc}[MPC]{multipath component}
\acrodef{fa}[FA]{false alarm}
\acrodef{far}[FAR]{false alarm rate}
\acrodef{pva}[PR]{potential ray}
\acrodef{mf}[MF]{map feature}
\acrodef{smc}[SMC]{specular multipath component}
\acrodef{psmc}[PSMC]{potential \ac{smc}}
\acrodef{mva}[SFV]{surface feature vector}
\acrodef{ip}[IP]{interaction point}
\acrodef{pmva}[PSFV]{potential \ac{mva}}
\acrodef{slam}[SLAM]{simultaneous localization and mapping}
\acrodef{Mp}[MP]{Multipath-based}
\acrodef{mpslam}[MP-SLAM]{multipath-based simultaneous localization and mapping}
\acrodef{pmf}[PMF]{probability mass function}
\acrodef{pdf}[PDF]{probability density function}
\acrodef{cdf}[CDF]{cumulative distribution function}
\acrodef{bp}[BP]{belief propagation}
\acrodef{spa}[SPA]{sum-product algorithm}
\acrodef{fg}[FG]{factor graph}
\acrodef{mmse}[MMSE]{minimum mean-square error}
\acrodef{simo}[SIMO]{single-input multiple-output}
\acrodef{miso}[MISO]{multiple-input single-output}
\acrodef{mimo}[MIMO]{multiple-input multiple-output}
\acrodef{dmimo}[D-MIMO]{distributed \ac{mimo}}
\acrodef{mmwave}[mmWave]{millimeter-wave}
\acrodef{ue}[UE]{mobile user}
\acrodef{los}[LoS]{line-of-sight}
\acrodef{nlos}[NLoS]{non \ac{los}}
\acrodef{olos}[OLoS]{obstructed \ac{los}}
\acrodef{aoa}[AoA]{angle-of-arrival}
\acrodef{aod}[AoD]{angle-of-departure}
\acrodef{roi}[RoI]{region-of-interest}
\acrodef{snr}[SNR]{signal-to-noise ratio}
\acrodef{ospa}[OSPA]{optimal subpattern assignment}
\acrodef{mospa}[MOSPA]{mean \ac{ospa}}
\acrodef{dmc}[DMC]{dense multipath component}
\acrodef{sr}[SR]{super-resolution}
\acrodef{sbl}[SBL]{sparse Bayesian learning}
\acrodef{sde}[SDE]{sequential detection and estimation}
\acrodef{kest}[KEST]{Kalman enhanced super resolution tracking}
\acrodef{da}[DA]{data association}
\acrodef{fim}[FIM]{Fisher information matrix}
\acrodef{crlb}[CRLB]{Cram\'{e}r–Rao lower bound}
\acrodef{pcrlb}[PCRLB]{posterior Cram\'{e}r–Rao lower bound}
\acrodef{meb}[MEB]{mapping error bound}
\acrodef{peb}[PEB]{position error bound}
\acrodef{oeb}[OEB]{orientation error bound}
\acrodef{va}[VA]{virtual anchor}
\acrodef{eadf}[EADF]{effective aperture distribution function}
\acrodef{lhf}[LHF]{likelihood function}
\acrodef{rmse}[RMSE]{root mean square error} 
\acrodef{ps}[PS]{point scatterer}
\acrodef{imm}[IMM]{interacting multiple model}
\acrodef{rf}[RF]{radio frequency}
\acrodef{iid}[iid]{independent and identically distributed}
\acrodef{prop}[AM-SLAM]{adaptive multipath-based SLAM algorithm}

\acrodef{refa}[MP-SLAM]{multipath-based SLAM}
\acrodef{chslam}[CH-SLAM]{channel SLAM}
\acrodef{rt}[RT]{ray tracing}
\acrodef{ceda}[CEDA]{snapshot-based parametric channel estimation and detection algorithm}

%% file: pgfplot.tex
\usetikzlibrary{backgrounds}
\usetikzlibrary{calc,arrows.meta}
\ifthenelse{\equal{\exportFigures}{true}}
{
  \usepgfplotslibrary{external}\tikzexternalize[prefix=\tikzfolder]
  \tikzset{external/system call={pdflatex \tikzexternalcheckshellescape -halt-on-error -interaction=batchmode -jobname "\image" "\texsource"}}

}
{}

\def\addlegendimage{\csname pgfplots@addlegendimage\endcsname}
\def\addlegendentry{\csname pgfplots@addlegendentry\endcsname}

\newcommand{\pgfref}[1]
{\ifthenelse{\equal{\exportFigures}{true}}
{\tikzexternaldisable\ref{#1}\tikzexternalenable}
{\ref{#1}}}

\pgfplotsset{compat=newest} 

\newlength\figureheight 
\newlength\figurewidth 
\usetikzlibrary{patterns,arrows}
\usepgfplotslibrary{colormaps}
\usetikzlibrary{plotmarks}
\pgfplotsset{every axis/.append style={
  label style={font=\normalsize},
  tick label style={font=\scriptsize},
  xticklabel={
   \ifdim \tick pt < 0pt
    \pgfmathparse{abs(\tick)}%
    \llap{$-{}$}\pgfmathprintnumber{\pgfmathresult}
   \else
    \pgfmathprintnumber{\tick}
   \fi}
   }}
\usetikzlibrary{decorations.markings}
\makeatletter
\tikzset{
  nomorepostactions/.code={\let\tikz@postactions=\pgfutil@empty},
  mymarkfixednumber/.style n args={3}{decoration={markings,
    mark= between positions 0.1 and 1 step (1/#3)*\pgfdecoratedpathlength with{%
        \tikzset{#2,every mark}\tikz@options
        \pgfuseplotmark{#1}%
      },  
    },
    postaction={decorate},
    /pgfplots/legend image post style={
        mark=#1,mark options={#2},every path/.append style={nomorepostactions}
    },
  },
}
\tikzset{
  nomorepostactions/.code={\let\tikz@postactions=\pgfutil@empty},
  mymarkfixeddistance/.style n args={3}{decoration={markings,
    mark= between positions 0 and 1 step #3cm with{%
        \tikzset{#2,every mark}\tikz@options
        \pgftransformresetnontranslations%
        \pgfuseplotmark{#1}%
      },  
    },
    postaction={decorate},
    /pgfplots/legend image post style={
        mark=#1,mark options={#2},every path/.append style={nomorepostactions}
    },
  },
}
\tikzset{
  nomorepostactions/.code={\let\tikz@postactions=\pgfutil@empty},
  mymark/.style n args={3}{decoration={markings,
    mark= between positions 0 and 1 step 0.75cm with{%
        \tikzset{#2,every mark}\tikz@options
        \pgftransformresetnontranslations%
        \pgfuseplotmark{#1}%
      },  
    },
    postaction={decorate},
    /pgfplots/legend image post style={
        mark=#1,mark options={#2},every path/.append style={nomorepostactions}
    },
  },
}
\makeatother

\definecolor{colA}{rgb}{0,0,1}
\definecolor{colB}{rgb}{1.00000,0.0,0.00000}
\definecolor{colC}{rgb}{0.1333,0.5451,0.1333}
\definecolor{colD}{rgb}{0.9,0.0,0.9}
\definecolor{colE}{rgb}{0,1,1}
\definecolor{colF}{rgb}{1.00000,0.55,0.00000}

\def\mymarksize{1.5}
\def\mymarksize2{2.5}

\pgfdeclareplotmark{markarrow}
{%
  \pgfpathmoveto{\pgfqpoint{\pgfplotmarksize}{0pt}}
  \pgfpathlineto{\pgfqpointpolar{130}{1.5\pgfplotmarksize}}
  \pgfpathmoveto{\pgfqpoint{\pgfplotmarksize}{0pt}}
  \pgfpathlineto{\pgfqpointpolar{-130}{1.5\pgfplotmarksize}}
  \pgfusepathqstroke
}

%% file: InputFiles/abstract.tex
Localizing users and mapping the environment using radio signals is a key task in emerging applications such as reliable low-latency communications, location-aware security, and safety-critical navigation. Recently introduced multipath-based \ac{slam} methods can jointly localize a mobile agent and map reflective surfaces in \ac{rf} environments. Most existing approaches assume that map features and their corresponding \ac{rf} propagation paths are statistically independent, conditioned on the state of the mobile agent. This assumption neglects inherent dependencies that arise when a single reflective surface contributes to multiple propagation paths or when an agent communicates with multiple base stations. Existing approaches that aim to fuse information across propagation paths are further limited by their inability to perform ray tracing in \ac{rf} environments with nonconvex geometries.

In this paper, we propose a Bayesian multipath-based \ac{slam} method for distributed \ac{mimo} systems that addresses these limitations. In particular, we exploit amplitude statistics to establish adaptive, time-varying detection probabilities. Based on the resulting ``soft'' ray-tracing strategy, the proposed method can fuse information across propagation paths in \ac{rf} environments with nonconvex geometries. A Bayesian estimation framework for the joint estimation of map features and agent state is developed by applying the message passing rules of the \ac{spa} to a factor graph representation of the proposed statistical model. We further introduce a new initialization procedure for reflective surfaces that enables the introduction of new surface states even when measurements arise solely from double-bounce paths. The proposed method is validated using both synthetic and real \ac{rf} measurements obtained in challenging scenarios with nonconvex geometries and obstructed line-of-sight conditions. The results demonstrate that it provides accurate localization and mapping performance and approaches the posterior Cram\'{e}r--Rao lower bound.

%% file: InputFiles/Introduction.tex
Ensuring reliable communication and sensing in harsh propagation environments, such as urban canyons and indoor spaces, remains a fundamental and challenging problem. Multipath-based \acf{slam} \cite{GentnerTWC2016, LiTWC2019, Erik_SLAM_TWC2019, RicoJSAC2019, YuGe_mmWaveSLAM_JSAC2022, MohammadmmWaveTVT2023, LeiVenTeaMey:TSP2023, LiLeiCaiTuf:ICC2024, LiangTSP2025} 
\begin{figure}[!t]
	\captionsetup[subfigure]{labelformat=empty, labelsep=none} 
	\centering
	\vspace*{-4mm}
	\hspace*{6mm}\subfloat[]
	{\hspace*{-6mm}\includegraphics[scale=0.98]{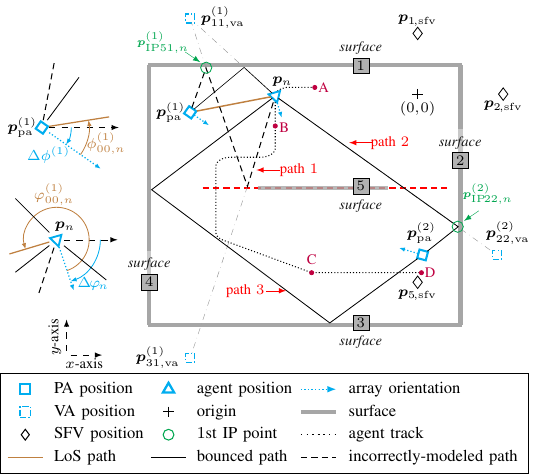}}\\[-4mm]
	\caption{Exemplary \ac{dmimo} \ac{rf} propagation scenario with five reflective surfaces $s \in \{1,\dots,5\}$, two \acp{pa} $\V{p}_{\mathrm{pa}}^{(j)}$, and one mobile agent $\V{p}_{n}$. Reflective surfaces are represented by \acp{mva} $\V{p}_{s,\mathrm{\sfv}}$, i.e., mirror images of a common origin $(0,0)$ with respect to the surfaces. Reflected paths are modeled by \acp{va} $\V{p}_{ss',\mathrm{va}}^{(j)}$. The intersection point of each path with the first reflective surface is denoted by $\V{p}_{\mathrm{\ip}ss',n}^{(j)}$. This intersection point determines the \ac{aod}. Array orientations, \ac{aod}, and \ac{aoa} at \ac{pa}~$1$ and the agent, respectively, are shown on the left side of the figure. The red dashed line indicates the infinite extent of surface~$5$ as assumed in previously proposed methods, e.g., \cite{LeiVenTeaMey:TSP2023}. In the absence of surface extent information, such an environmental model may introduce nonexistent paths such as path~$1$, while failing to account for existing paths such as paths~$2$ and~$3$, since the infinite extent of surface~$5$ would block these paths.}
	\label{fig:GraphicalOverview_MVADemo}
	\vspace*{-4mm}
\end{figure}
in 5G and 6G networks has therefore attracted significant attention as a promising solution. Existing multipath-based \ac{slam} methods are typically categorized as feature-based \ac{slam}, in which propagation paths arising from \ac{rf} signals interacting with environmental objects are modeled individually using independent map features, such as \acp{va} and point scatterers. The \Ac{va} is a commonly used feature type representing mirror images of \acp{pa} (e.g., \acp{bs}) with respect to planar surfaces, thereby modeling specular \ac{rf} reflections. In general, the number of features and their positions are unknown and time-varying. Multipath-based \ac{slam} jointly estimates the mobile agent state and the locations of distinct features, either using measurements of \acp{mpc} that are extracted from received \ac{rf} signals via parametric channel estimation \cite{RichterPhD2005, ShutinCSTA2013, BadiuTSP2017, HansenTSP2018, grebien2024SBL, MoeWesVenLei:Fusion2025}, or directly from \ac{rf} signals \cite{LiangTSP2025}.

\vspace*{8mm}

\subsection{State-of-the-Art Methods}
A common assumption in state-of-the-art multipath-based \ac{slam} methods is that \acp{va} corresponding to individual propagation paths are conditionally independent given the state of the agent. This assumption is restrictive, as multiple propagation paths may originate from interactions with the same environmental feature. In particular, a reflective surface (e.g.,~a wall) is an extended object that can simultaneously generate single-bounce as well as higher-order multi-bounce paths, all of which may significantly contribute to the received signal.

Recent work addresses this limitation by considering the mapping of extended objects \cite{MeyerWilliams:TSP2021} and by exploiting data fusion across \acp{mpc} in multipath-based \ac{slam} \cite{Hyowon2020TWC, Chu2022TWC, YanWenJinLi:TWC2022, LeiVenTeaMey:TSP2023, WieVenWilWitLei:Fusion2024, GeWymSev:TSP2025}, demonstrating notable performance improvements. In particular, \cite{Hyowon2020TWC, Chu2022TWC, LeiWieVenWit:Asilomar2024} propose methods that fuse information across single-bounce paths and cooperating agents. Furthermore, \cite{WieVenWilLei:JAIF2023, WieVenWilWitLei:Fusion2024} introduce statistical models that jointly account for diffuse \acp{mpc} arising from rough or non-ideal reflective surfaces, while \cite{LeiVenTeaMey:TSP2023} develops the master virtual anchors (MVAs) model for representing reflective surfaces that enables data fusion across paths up to double-bounce reflections and \acp{pa}.

A simple geometric model for a reflective surface, as used in \cite{LeiVenTeaMey:TSP2023}, represents the surface as an infinite straight line characterized by its position and orientation, as well as binary decisions on the availability of propagation paths obtained via conventional ``hard'' ray tracing. While this model has been demonstrated to be effective in certain indoor experiments, it is essentially limited to convex room geometries. This is because in scenarios with nonconvex geometries, the infinite nature of the reflective surface represented by the \acp{mva} ``blocks'' physically possible propagation path in the hard ray tracing process, making it impossible to reliably decide on the availability of propagation paths. An extension to nonconvex room geometries requires either (i) modeling and estimating the shape and size of the surface represented by the MVA; (ii) inferring the time-varying availability of propagation paths associated with each MVA. Approach (i) would introduce significant additional complexity and require dense measurements across the surface to yield reliable extent estimates \cite{Chu2022TWC, WieVenWilWitLei:Fusion2024}. For this reason, we focus on approach (ii) in this paper.

State-of-the-art methods for localization and tracking using \acp{mpc} \cite{XuhongTWC2022, VenusTWC2024} perform the estimation of complex path amplitudes that determine the detection probability of propagation paths and measurement variances related to each surface within a joint sequential Bayesian framework. The joint formulation enables reliable and adaptive feature detection even in the presence of measurement impairments such as clutter and missed detections.

\subsection{Contributions}

This paper introduces an \ac{prop} for \acf{dmimo} systems that jointly estimates the position and orientation of the mobile agent, detects and localizes reflective surfaces, and is applicable to arbitrary room geometries.

A key novelty of the proposed approach is the introduction of a normalized amplitude and a binary existence variable for each potential first- and second-order propagation path, enabling the inference of the time-varying availability and strength of each path within a unified Bayesian framework. The method operates on \ac{mpc} parameter estimates, e.g., distances, \acp{aoa}, \acp{aod}, and normalized amplitudes, obtained in a pre-processing stage. Reflective surfaces are modeled by \acfp{mva}, formerly referred to as MVAs \cite{LeiVenTeaMey:TSP2023}\footnote{We adopt the term \ac{mva} instead of MVA for improved generality, as the proposed model can accommodate additional surface parameters, such as spatial extent or other properties describing the surface.}, which characterize their positions and orientations. Based on the \acp{mva}, the corresponding \acp{va} can be derived to model either single- or double-bounce propagation paths via geometric transformations.

We introduce a new statistical model with hierarchical features, in which \ac{mva} states reside at the upper level of the hierarchy and propagation paths at the lower level. Each \ac{mva} gives rise to multiple propagation paths. As in our previous work \cite{XuhongTWC2022, VenusTWC2024}, the states of the propagation paths are their complex amplitudes, which determine both the detection probability of the paths and the associated measurement variances. Both \ac{mva} states and path amplitude states are augmented with existence variables that enable inference of their existence probabilities. The proposed hierarchical feature model facilitates estimation of the number of \acp{mva} and enables ``soft'' ray tracing. In particular, by explicitly modeling propagation paths and inferring their existence, the method can capture nonconvex room geometries. Another key aspect is that new \acp{mva} can be initialized even when the corresponding reflective surface yields measurements only from double-bounce paths. This capability contrasts with \cite{LeiVenTeaMey:TSP2023} and is enabled by an improved initialization procedure for new \acp{mva}.

The proposed method performs probabilistic \ac{da} and sequential estimation of the states of the mobile agent, \acp{mva}, and \acp{va} by running \ac{spa} on a \ac{fg} representation of the proposed statistical model. Numerical results, based on both synthetic and real data, compare the proposed method with state-of-the-art reference methods \cite{GentnerTWC2016, Erik_SLAM_TWC2019, LeitingerICC2019} as well as the \acp{pcrlb}. The results demonstrate that the proposed method significantly improves both localization and mapping accuracy, achieving state-of-the-art performance. The key contributions of this paper can be summarized as follows\vspace{0mm}.
\begin{itemize}
	\item We introduce a statistical model for multipath-based SLAM based on \acp{mva} that can represent the presence and strength of each propagation path.
	\vspace{2mm}
	\item We develop the \ac{fg} and a \ac{spa} for multipath-based \ac{slam} D-MIMO systems with adaptive data fusion across propagation paths suitable for nonconvex room geometries\vspace{2mm}.
	\item We design a proposal \ac{pdf} that relies on both single- and double-bounce propagation paths, enabling effective introduction of new surfaces.
	\vspace{2mm}
	\item We validate our method using synthetic and real measurements and compare performance against two state-of-the-art methods \cite{GentnerTWC2016, Erik_SLAM_TWC2019, LeitingerICC2019} and the \acp{pcrlb}\vspace{0mm}.
\end{itemize}

This work advances over the previous work on multipath-based \ac{slam} exploiting a \ac{mva} model \cite{LeiVenTeaMey:TSP2023} by (i) incorporating amplitude statistics into the joint statistical model for adaptive path detection and fusion; (ii) extending to a \ac{dmimo} setup; (iii) improving the proposal \ac{pdf} for new \acp{mva}; (iv) providing the \acp{pcrlb} as a performance benchmark; and (v) validating the performance using both synthetic and real measurements in challenging dynamic scenarios. Note that the detailed derivations and the necessary references for the Bayesian models and the corresponding \ac{spa} are provided in the supplementary material \cite{MVASLAM_TWC2025SupportingDoc} to ensure clarity and completeness of the methodology.

\textit{Notations}: Column vectors and matrices are denoted as lowercase and uppercase bold letters. Random variables are displayed in sans serif, upright fonts as for example $\rv{x}$ and $\RV{x}$ and their realizations in serif, italic font as for example $x$ and $\V{x}$. $f(\V{x})$ denotes the \ac{pdf} or \ac{pmf} of a continuous or discrete random vector. $(\cdot)^{\mathrm{T}}$ denotes matrix transpose. $ \norm{\cdot} $ is the Euclidean norm. $ \vert\cdot\vert $ represents the cardinality of a set. $f(\V{a}|\V{b})$ denotes the conditional \ac{pdf} of random vector $\RV{a}$ conditioned on random vector $\RV{b}$. $f_{\mathrm{N}}(z|a,b)$ denotes Gaussian \ac{pdf} with mean $a$ and variance $b^2$. $ \angle(\V{a}, \V{b}, o) \triangleq \mathrm{atan2}\big(\V{a}, \V{b}\big) - o $ calculates the angle between the two points $\V{a}$ and $\V{b}$ w.r.t. the reference angle $o$.

%% file: InputFiles/GeometricalRelations.tex
We consider a \ac{dmimo} system operating in a dynamic scenario and consisting of $ J $ distributed static base stations (i.e., \acp{pa}) and a mobile agent, as shown in Fig.\,\ref{fig:GraphicalOverview_MVADemo}. 
For the sake of brevity, we assume a \ac{2d} scenario with horizontal-only \ac{rf} signal propagation. At each discrete time $ n $, all $ J $ \acp{pa} with known positions $\V{p}_{\mathrm{pa}}^{(j)} =[p_{\mathrm{pa},\mathrm{x}}^{(j)} \iist p_{\mathrm{pa},\mathrm{y}}^{(j)}]^{\mathrm{T}} \rmv\in\rmv \mathbb{R}^{2\rmv\times\rmv1} $ and $j \in \{1,\dots,J\}$ transmit \ac{rf} signals and the mobile agent at an unknown, time-varying position $\V{p}_{n} =[p_{n,\mathrm{x}} \iist p_{n,\mathrm{y}}]^{\mathrm{T}} \rmv\in\rmv \mathbb{R}^{2\rmv\times\rmv1} $ acts as a receiver. Time and frequency synchronization between the \acp{pa} and the agent are assumed to be perfect.\footnote{The proposed algorithm can be easily extended to \ac{3d} scenarios with both horizontal and vertical propagation and reformulated for the case where the mobile agent acts as a transmitter and the \acp{pa} act as receivers, as shown in Section~\ref{subsec:RealResults}. It can also be extended to unsynchronized systems in line with
\cite{GentnerTWC2016, KimGranSveKimWym:TVT2022, FasDeuKesWilColWitLeiSecWym:STSP2025, Junshi_arXiv2025}.} Each \ac{pa} is equipped with a $ H_{\mathrm{tx}} $-element antenna array with known orientation $\Delta\phi^{(j)}$, and the mobile agent is equipped with a $ H_{\mathrm{rx}} $-element antenna array with unknown orientation $\Delta\varphi_{n}$. The positions $ \V{p}_{\mathrm{pa}}^{(j)} $ and $ \V{p}_{n} $ refer to the center of gravity of the arrays.

\subsection{SFV-Based Model of the Environment}
\label{sec:MVAEnvModel}
The \ac{rf} signal propagation environment consists of $ S $ static reflective surfaces which are indexed by $s \in \mathcal{S} \triangleq \{1,\dots,S\} $ and modeled by \acp{mva} at positions $ \V{p}_{s,\mathrm{\sfv}} \rmv\in\rmv \mathbb{R}^{2\rmv\times\rmv1}$ denoting the mirror images of the origin $[0,0]^\mathrm{T} $ on the surfaces. Propagation paths that are specularly reflected can be modeled by \acp{va} at positions $ \V{p}_{ss',\mathrm{va}}^{(j)} \rmv\in\rmv \mathbb{R}^{2\rmv\times\rmv 1} $ denoting the mirror images of the $j$th \ac{pa} on these surfaces, where the index tuple $ (s\rmv,\rmv s') \in \mathcal{D} \triangleq \{(s\rmv,\rmv s') \in \mathcal{S}\rmv\rmv\times\rmv\rmv\mathcal{S} \} $ denotes the signal interaction order from the $ s'$th surface to the $ s$th surface before reaching the agent. Following \cite{LeiVenTeaMey:TSP2023}, we consider paths experiencing maximal double bounces\footnote{Propagation paths up to second-order reflections constitute the primary contribution to the received signals, making this a practical assumption. However, the model can be extended to incorporate higher-order bounce paths, though this introduces a substantial increase in complexity.}. The two sets $ \mathcal{D}_{\mathrm{S}} \triangleq \{(s\rmv,\rmv s) \in \mathcal{S}\rmv\rmv\times\rmv\rmv\mathcal{S} \} $ with $ \vert \mathcal{D}_{\mathrm{s}}\vert = S $ and $ \mathcal{D}_{\mathrm{D}} \triangleq \{(s\rmv,\rmv s') \in \mathcal{S}\rmv\rmv\times\rmv\rmv\mathcal{S} | s\neq s'\} $ with $ \vert \mathcal{D}_{\mathrm{D}}\vert = S(S-1) $ contain the index tuples for single-bounce and double-bounce \acp{va}, respectively. Therefore, the maximum number of \acp{va} for \ac{pa} $j$ is given by $ \vert \mathcal{D}\vert = S+S(S-1) $ and $\mathcal{D} = \mathcal{D}_{\mathrm{S}} \cup \mathcal{D}_{\mathrm{D}}$. For conveniently addressing \ac{pa}-related notations, we also define $ \V{p}_{\mathrm{pa}}^{(j)} = \V{p}_{00,\mathrm{va}}^{(j)} $ and $\tilde{\mathcal{D}} = (0,\rmv 0)\rmv \cup \rmv \mathcal{D}$. The positions of a single-bounce \ac{va} $\V{p}_{ss,\mathrm{va}}^{(j)}$ and a double-bounce \ac{va} $\V{p}_{ss',\mathrm{va}}^{(j)}$ can be transformed from $\V{p}_{s,\mathrm{\sfv}}$, $\V{p}_{s',\mathrm{\sfv}}$ and $\V{p}_{00,\mathrm{va}}^{(j)}$ as $\V{p}_{ss,\mathrm{va}}^{(j)} = h_{\mathrm{va}} \big( \V{p}_{s,\mathrm{\sfv}}, \V{p}_{\mathrm{pa}}^{(j)}\big)$ and $\V{p}_{ss',\mathrm{va}}^{(j)} = h_{\mathrm{va}} \big( \V{p}_{s,\mathrm{\sfv}}, h_{\mathrm{va}} \big( \V{p}_{s',\mathrm{\sfv}}, \V{p}_{\mathrm{pa}}^{(j)}\big) \big)$, respectively. The inverse transformation from a \ac{va} position to a \ac{mva} position is given by $\V{p}_{s,\mathrm{\sfv}}^{(j)} = h_{\mathrm{\sfv}} \big( \V{p}_{ss,\mathrm{va}}^{(j)} , \V{p}_{\mathrm{pa}}^{(j)}\big)$. Details of the transformations $h_{\mathrm{va}} \big( \cdot, \cdot\big)$ and $h_{\mathrm{\sfv}} \big( \cdot, \cdot\big)$ are provided in \cite[Section~II-A]{LeiVenTeaMey:TSP2023}.

For each propagation path, its distance $d_{ss',n}^{(j)}$, \ac{aod} $\phi_{ss',n}^{(j)}$ and \ac{aoa} $\varphi_{ss',n}^{(j)}$ are modeled by $ d_{ss',n}^{(j)} = \|\V{p}_{n} - \V{p}_{ss',\mathrm{va}}^{(j)}\| $, $ \phi_{ss',n}^{(j)} = \angle (\V{p}_{\mathrm{\ip}ss',n}^{(j)}, \V{p}_{\mathrm{pa}}, \Delta\phi^{(j)})$ and $ \varphi_{ss',n}^{(j)} = \angle (\V{p}_{n}, \V{p}_{ss',\mathrm{va}}^{(j)}, \Delta\varphi_{n})$. The \ac{ip} $\V{p}_{\mathrm{\ip}ss',n}^{(j)}$ of path on the first interacting surface $s'$ is given by 
\begin{align}
	\V{p}_{\mathrm{\ip}ss',n}^{(j)} = \V{p}_{\mathrm{pa}}^{(j)}  + \frac{(\V{p}_{s'}^{\mathrm{w}} - \V{p}_{\mathrm{pa}}^{(j)} )^\mathrm{T} \V{n}_{s'}^{\mathrm{w}} }{( \V{p}_{ss',n}^{\mathrm{vm}} - \V{p}_{\mathrm{pa}}^{(j)} )^\mathrm{T} \V{n}_{s'}^{\mathrm{w}} } (\V{p}_{ss',n}^{\mathrm{vm}} - \V{p}_{\mathrm{pa}}^{(j)} )
	\label{eq:BouncPoint2B} 
\end{align}
where $\V{p}_{s'}^{\mathrm{w}} = \frac{\V{p}_{s',\mathrm{\sfv}}}{2}$, $\V{n}_{s'}^{\mathrm{w}} = \frac{\V{p}_{s',\mathrm{\sfv}}}{\| \V{p}_{s',\mathrm{\sfv}} \|}$ is the normal vector of the $s'$th reflective surface, and the virtual mobile agent positions $\V{p}_{ss',n}^{\mathrm{vm}} $ are computed by applying the same transformations for obtaining \acp{va} from \acp{mva}, leading to $ \V{p}_{ss,n}^{\mathrm{vm}} = h_{\mathrm{va}} \big( \V{p}_{s,\mathrm{\sfv}}, \V{p}_{n}\big)$ for single-bounce paths ($(s\rmv,\rmv s') \in \mathcal{D}_{\mathrm{S}}$), and $ \V{p}_{ss',n}^{\mathrm{vm}} = h_{\mathrm{va}} \big( \V{p}_{s,\mathrm{\sfv}}, h_{\mathrm{va}} \big( \V{p}_{s',\mathrm{\sfv}}, \V{p}_{n} \big) \big)$ for double-bounce paths ($(s\rmv,\rmv s') \in \mathcal{D}_{\mathrm{D}}$). For the $j$th \ac{pa} related \ac{los} path, its parameters are defined as $d_{00,n}^{(j)}$, $\phi_{00,n}^{(j)}= \angle (\V{p}_{n}, \V{p}_{\mathrm{pa}}^{(j)}, \Delta\phi^{(j)})$ and $\varphi_{00,n}^{(j)}$.

%% file: InputFiles/SystemModel.tex
At each time $ n $, the state of the mobile agent is given by $ \RV{x}_{n} \triangleq [\RV{p}_{n}^{\mathrm{T}} \iist \RV{v}_{n}^{\mathrm{T}} \iist \rv{{\Delta\varphi}}_{n}]^{\mathrm{T}}$ consisting of the position $ \RV{p}_{n} $, the velocity $ \RV{v}_{n} =[\rv{v}_{\mathrm{x},n} \iist \rv{v}_{\mathrm{y},n}]^{\mathrm{T}} $ and the azimuth array orientation $ \rv{{\Delta\varphi}}_{n}$. The agent states for all time steps up to $ n $ are denoted as $ \RV{x}_{1:n} \triangleq [\RV{x}_{1}^{\mathrm{T}} \ist\cdots\ist \RV{x}_{n}^{\mathrm{T}} ]^{\mathrm{T}} $. Following \cite{LeiVenTeaMey:TSP2023}, we account for the unknown number of \acp{mva} at each time step $n$ by introducing \acp{pmva} indexed by $s \in \{1,\dots, S_{n}\}$, where $ S_{n} $ represents the maximum possible number of \acp{pmva} that have been produced at least one measurement so far and $ S_{n} $ increases with time. \Ac{pmva} states are given as $ \RV{y}_{s,n} \triangleq [\RV{p}_{s,\mathrm{\sfv}}^{\mathrm{T}} \iist \rv{r}_{s,n}]^{\mathrm{T}}$ with $\RV{p}_{s,\mathrm{\sfv}}$ denoting the position. The existence/nonexistence of the $s$th \ac{pmva} is modeled by a binary random variable $\rv{r}_{s,n} \in \{1, 0\} $ in the sense that it exists if and only if $r_{s,n} = 1$. Similarly, to account for the unknown and time-varying number of \acp{va} associated to each \ac{pmva}, we introduce for each \ac{va} a single \ac{pva} with state $ \RV{{\beta}}_{ss',n}^{(j)} \triangleq [ \rv{{u}}_{ss',n}^{(j)} \iist \rv{r}_{ss',n}^{(j)}]^{\mathrm{T}}$ consisting of the normalized amplitude $\rv{u}_{ss',n}^{(j)}$ and the binary existence variable $\rv{r}_{ss',n}^{(j)} \in \{1, 0\} $. 

We further define the state of the $j$th \ac{pa} as $ \RV{{\beta}}_{00,n}^{(j)} \triangleq [ \rv{{u}}_{00,n}^{(j)} \iist \rv{r}_{00,n}^{(j)}]^{\mathrm{T}}$ to account for its time-varying \ac{los} condition to the agent. Fundamentally, a \ac{pmva} models a potentially existing environmental object and \acp{pva} model signal interacting processes on this object, thus their existence/nonexistence status are related as follows: (i) a nonexistent reflective surface cannot generate any propagation paths, i.e., from $r_{s,n} = 0$ follows $r_{ss',n}^{(j)} = 0 \, \forall (s\rmv,\rmv s')$; (ii) an existing surface does not necessarily interact with any \ac{rf} signals and therefore generate propagation paths, i.e., if $r_{s,n} = 1$, then $r_{ss',n}^{(j)} \in \{1,0\}$; (iii) a path exists if and only if the interacting surfaces all exist, i.e., from $r_{ss',n}^{(j)} = 1$ follows $r_{s,n} = 1$ and $r_{s',n} = 1$. The aforementioned relations are considered in the pseudo \ac{lhf} formulations in Section \ref{sec:jointPDFandFG}.

Formally, the states of nonexistent \acp{pmva} and nonexistent \acp{pva}, i.e., $r_{s,n} = 0$ and $r_{ss',n}  = 0$, are also considered even though they are irrelevant and have no influence on the detection and state estimation. Therefore, all \acp{pdf} defined for \ac{pmva} and \ac{pva} states $ f(\V{y}_{s,n}) = f(\V{p}_{s,\mathrm{\sfv}}, r_{s,n}) $ and $ f(\V{{\beta}}_{ss',\mathrm{va}}^{(j)}) = f(u_{ss',n}^{(j)}, r_{ss',n}^{(j)}) $, are of the forms $ f(\V{{p}}_{s,\mathrm{\sfv}}, 0 ) = f_{s,n} f_{\mathrm{D}}(\V{y}_{s,n}) $ and $ f({u}_{ss',n}^{(j)}, 0) = f_{ss',n}^{(j)} f_{\mathrm{D}}(u_{ss',n}^{(j)}) $, respectively, where $ f_{\mathrm{D}}(\cdot) $ is an arbitrary ``dummy \ac{pdf}'', and $ f_{s,n} \in [0,1]$ and $ f_{ss',n}^{(j)} \in [0,1]$ are constants representing the probabilities of nonexistence \cite{Florian_Proceeding2018, Florian_TSP2017, Erik_SLAM_TWC2019}.

\subsection{Measurements and New \acp{pmva}}
\label{subsec:MeaModel}

The mobile agent state, \ac{pmva} states and \ac{pva} states relate to the distance measurements ${\rv{z}_\mathrm{d}}_{m,n}^{(j)}$, the \ac{aod} measurements $ {\rv{z}_\mathrm{\phi}}_{m,n}^{(j)} $, the \ac{aoa} measurements ${\rv{z}_\mathrm{\varphi}}_{m,n}^{(j)}$, and the normalized amplitude measurements ${\rv{z}_\mathrm{u}}_{m,n}^{(j)}\rmv\in\rmv [u_{\mathrm{de}}, \infty) $ via the following \acp{lhf}, which are assumed to be conditionally independent of each other. The individual \acp{lhf} of the distance, \ac{aod} and \ac{aoa} measurements are modeled by Gaussian \acp{pdf}, given as 
\begin{align}
	&f_{ss'}^{(j)}({z_\mathrm{d}}_{m,n}^{(j)}) \triangleq f_{\mathrm{N}}({z_\mathrm{d}}_{m,n}^{(j)}; d_{ss',n}^{(j)}, \sigma_{\mathrm{d}}(u_{ss',n}^{(j)})), 
	\label{eq:LHF_dist}  \vspace*{2mm}\\
	&f_{ss'}^{(j)}({z_\mathrm{\phi}}_{m,n}^{(j)}) \triangleq f_{\mathrm{N}}({z_\mathrm{\phi}}_{m,n}^{(j)} ; \phi_{ss',n}^{(j)}, \sigma_{\mathrm{\phi}}(u_{ss',n}^{(j)})),  
	\label{eq:LHF_AoD} \vspace*{2mm}\\
	&f_{ss'}^{(j)}({z_\mathrm{\varphi}}_{m,n}^{(j)}) \triangleq f_{\mathrm{N}}({z_\mathrm{\varphi}}_{m,n}^{(j)} ; \varphi_{ss',n}^{(j)}, \sigma_{\mathrm{\varphi}}(u_{ss',n}^{(j)})) 
	\label{eq:LHF_AoA}
\end{align}
where the means $d_{ss',n}^{(j)}$, $\phi_{ss',n}^{(j)}$ and $\varphi_{ss',n}^{(j)}$ are calculated according to Section~\ref{sec:MVAEnvModel}. The variances of the Gaussian \acp{pdf} $ {\sigma_\mathrm{d}}^2(u_{ss',n}^{(j)}) $, $ \sigma_{\mathrm{\phi}}^2(u_{ss',n}^{(j)}) $ and $ \sigma_{\mathrm{\varphi}}^2(u_{ss',n}^{(j)}) $ are determined based on their respective Fisher information as in \cite{Thomas_Asilomar2018, LeitingerICC2019}, i.e., $ {\sigma_\mathrm{d}}^2(u_{ss',n}^{(j)}) = c^2/(8\pi^2 \beta_{\mathrm{bw}}^2 (u_{ss',n}^{(j)})^2) $, $ \sigma_{\phi}^2(u_{ss',n}^{(j)})  = c^2/(8\pi^2f_{\mathrm{c}}^2 (u_{ss',n}^{(j)})^2 D^2(\phi_{ss',n}^{(j)})) $ and $ \sigma_{\varphi}^2(u_{ss',n}^{(j)}) = c^2/(8\pi^2f_{\mathrm{c}}^2 (u_{ss',n}^{(j)})^2 D^2(\varphi_{ss',n}^{(j)})) $ with $ \beta_{\mathrm{bw}}^2$ denoting the mean square bandwidth of the transmit signal pulse, $ D^2(\cdot) $ is the squared array aperture, $c$ is the speed of light and $f_{\mathrm{c}}$ is the center carrier frequency. The normalized amplitude $u_{ss',n}^{(j)}$ is defined as the square root of the propagation path's \ac{snr}, and directly related to the detection probability $ {p_{\mathrm{d}}}(u_{ss',n}^{(j)}) $ of \ac{pva} $(s,s')$ via the Marcum Q-function as $ {p_{\mathrm{d}}}(u_{ss',n}^{(j)}) = Q_{1}(u_{ss',n}^{(j)}/\sigma_{\mathrm{u}}(u_{ss',n}^{(j)}), u_{\mathrm{de}}/\sigma_{\mathrm{u}}(u_{ss',n}^{(j)})) $ \cite{BarShalom_AlgorithmHandbook, XuhongTWC2022}. The scale parameter $ \sigma_{\mathrm{u}}(u_{ss',n}^{(j)}) $ is set to the corresponding Fisher information \cite{XuhongTWC2022}, i.e., $ \sigma_{\mathrm{u}}(u_{ss',n}^{(j)}) = \frac{1}{2} + \frac{1}{4N_{\mathrm{rx}}N_{\mathrm{s}}} (u_{ss',n}^{(j)})^2 $, with $N_{\mathrm{s}}$ denoting the number of signal samples per transmit--receive antenna pair at time $n$. The \acp{lhf} of the normalized amplitude measurements $ {z_\mathrm{u}}_{m,n}^{(j)} \rmv>\rmv u_{\mathrm{de}} $ are modeled by a truncated Rician \ac{pdf} \cite[Ch.\,1.6.7]{BarShalom_AlgorithmHandbook}\cite{XuhongTWC2022}, i.e.,
\begin{align}
	f_{ss'}^{(j)}({z_\mathrm{u}}_{m,n}^{(j)}) \triangleq  f_{\mathrm{TR}}({z_\mathrm{u}}_{m,n}^{(j)}; u_{ss',n}^{(j)}, \sigma_{\mathrm{u}}(u_{ss',n}^{(j)}), {p_{\mathrm{d}}}(u_{ss',n}^{(j)}); u_{\mathrm{de}}).
	\label{eq:LHF_normAmp} 
\end{align}
Note that \eqref{eq:LHF_dist}-\eqref{eq:LHF_normAmp} generally apply for \ac{los} paths, single- and double-bounce paths, depending on whether $(s\rmv,\rmv s') = (0,0)$, $(s\rmv,\rmv s') \in \mathcal{D}_{\mathrm{S},n}^{(j)}$ or $(s\rmv,\rmv s') \in \mathcal{D}_{\mathrm{D},n}^{(j)}$. Using \eqref{eq:LHF_dist} through \eqref{eq:LHF_normAmp}, the \acp{lhf} for different types of paths are factorized as follows and further used in the pseudo \acp{lhf} in Section~\ref{sec:jointPDFandFG}.
\subsubsection{\ac{lhf} for \ac{los} Paths} 
\begin{align}
	& f_{\mathrm{P}}(\V{z}_{m,n}^{(j)}|\V{p}_{n}, u_{00,n}^{(j)})  \nn \\ 
	& \hspace*{5mm} =  f_{00}^{(j)}({z_\mathrm{d}}_{m,n}^{(j)}) f_{00}^{(j)}({z_\mathrm{\phi}}_{m,n}^{(j)}) f_{00}^{(j)}({z_\mathrm{\varphi}}_{m,n}^{(j)}) f_{00}^{(j)}({z_\mathrm{u}}_{m,n}^{(j)}). 
	\label{eq:LHF_PA}
\end{align}
\subsubsection{\ac{lhf} for Single-Bounce Paths} 
\begin{align}
	& f_{\mathrm{S}}(\V{z}_{m,n}^{(j)}|\V{p}_{n}, \V{p}_{s,\mathrm{\sfv}}^{(j)}, u_{ss,n}^{(j)}) \nn \\ 
	& \hspace*{5mm} =  f_{ss}^{(j)}({z_\mathrm{d}}_{m,n}^{(j)}) f_{ss}^{(j)}({z_\mathrm{\phi}}_{m,n}^{(j)}) f_{ss}^{(j)}({z_\mathrm{\varphi}}_{m,n}^{(j)}) f_{ss}^{(j)}({z_\mathrm{u}}_{m,n}^{(j)}). 
	\label{eq:LHF_Spath}
\end{align}
\subsubsection{\ac{lhf} for Double-Bounce Paths} 
\begin{align}
	& f_{\mathrm{D}}(\V{z}_{m,n}^{(j)}|\V{p}_{n}, \V{p}_{s,\mathrm{\sfv}}^{(j)}, \V{p}_{s',\mathrm{\sfv}}^{(j)}, u_{ss',n}^{(j)})  \nn \\ 
	& \hspace*{5mm} =  f_{ss'}^{(j)}({z_\mathrm{d}}_{m,n}^{(j)}) f_{ss'}^{(j)}({z_\mathrm{\phi}}_{m,n}^{(j)}) f_{ss'}^{(j)}({z_\mathrm{\varphi}}_{m,n}^{(j)}) f_{ss'}^{(j)}({z_\mathrm{u}}_{m,n}^{(j)}).
	\label{eq:LHF_Dpath}
\end{align}
Before being observed, the measurements $ \RV{z}_{m,n}^{(j)} \rmv\triangleq\rmv [ {\rv{z}_\mathrm{d}}_{m,n}^{(j)} \iist {\rv{z}_\mathrm{\phi}}_{m,n}^{(j)} \iist {\rv{z}_\mathrm{\varphi}}_{m,n}^{(j)} \iist {\rv{z}_\mathrm{u}}_{m,n}^{(j)} ]^{\mathrm{T}} \rmv\in\rmv \mathbb{R}^{4\rmv\times\rmv1} $ with $m \in \{1,\dots,\rv{M}_{n}^{(j)}\}$ and the measurement number $\rv{M}_{n}^{(j)}$ at each \ac{pa} are considered as random. The joint measurement vectors for all \acp{pa} and all times up to $n$ are given by $ \RV{z}_{n}^{(j)} \rmv\triangleq\rmv [\RV{z}_{1,n}^{{(j)}\mathrm{T}} \ist \cdots \ist \RV{z}_{\rv{M}_{n}^{(j)},n}^{^{(j)}\mathrm{T}}]^{\mathrm{T}} \rmv\in\rmv \mathbb{R}^{4\rv{M}_{n}^{(j)}\rmv\times\rmv1}$, $ \RV{z}_{n} \rmv\triangleq\rmv [\RV{z}_{n}^{{(1)}\mathrm{T}} \ist \cdots \ist \RV{z}_{n}^{{(J)}\mathrm{T}}]^{\mathrm{T}} $ and $ \RV{z}_{1:n} \rmv\triangleq\rmv [\RV{z}_{1}^{\mathrm{T}} \ist \cdots \ist \RV{z}_{n}^{\mathrm{T}}]^{\mathrm{T}} $. Note that the measurements are obtained by applying a parametric channel estimation and detection algorithm \cite{RichterPhD2005, ShutinCSTA2013,BadiuTSP2017,HansenTSP2018, grebien2024SBL,MoePerWitLei:TSP2024,MoeWesVenLei:Fusion2025} to the observed discrete \ac{rf} signals in the pre-processing stage, where a detection threshold $ u_{\mathrm{de}} $ is applied to the normalized amplitude, leading to ${z_\mathrm{u}}_{m,n}^{(j)} > u_{\mathrm{de}} $. The measurements $\RV{z}_{m,n}^{(j)}$ that do not originate from any \acp{mva} are referred to as \acp{fa}, which are assumed to be statistically independent of \ac{pmva} and \ac{pva} states and modeled by a Poisson point process with mean number $ \mu_{\mathrm{fa}} $ and \acp{lhf} $ f_{\mathrm{fa}}(\V{z}_{m,n}^{(j)}) $.

\subsubsection{New \acp{pmva} and New \acp{pva}}
Newly detected \acp{mva}, i.e., \acp{mva} that generate measurements for the first time, are modeled by a Poisson point process with mean $\mu_{\mathrm{n}}$ and conditional \ac{pdf} $f_{\mathrm{n}}(\overline{\V{p}}_{m,\mathrm{\sfv}}^{(j)}| \V{x}_{n})$. Newly detected \acp{mva} are represented by new \ac{pmva} states $ \overline{\RV{y}}_{m,n}^{(j)} \triangleq [\overline{\RV{p}}_{m,\mathrm{\sfv}}^{{(j)}\mathrm{T}}\iist \overline{\rv{r}}_{m,n}^{(j)}]^{\mathrm{T}} $, $ m \in \{1,\dots, \rv{M}_n^{(j)}\} $. Each new \ac{pmva} $ \overline{\RV{y}}_{m,n}^{(j)} $ corresponds to a new \ac{pva} $\overline{\RV{{\beta}}}_{m,n}^{(j)} = [\overline{\rv{u}}_{mm,n}^{(j)}, \overline{\rv{r}}_{mm,n}^{(j)} ]^{\mathrm{T}}$ and a measurement $\RV{z}_{m,n}^{(j)}$. If and only if $\overline{r}_{m,n}^{(j)}=1$ and $\overline{r}_{mm,n}^{(j)} = 1$, the measurement $\RV{z}_{m,n}$ is claimed to be generated from the new \ac{pmva} $m$. The normalized amplitude states of new \acp{pva} are modeled by the \ac{pdf} $f_{\mathrm{n}}(\overline{u}_{mm,n}^{(j)})$. The state vectors of all new \acp{pmva} and new \acp{pva} at \ac{pa} $j$ are given by $ \overline{\RV{y}}_{n}^{(j)} \triangleq [\overline{\RV{y}}_{1,n}^{{(j)}\mathrm{T}} \ist\cdots\ist \overline{\RV{y}}_{\rv{M}_{n}^{(j)},n}^{{(j)}\mathrm{T}} ]^{\mathrm{T}} $ and $ \overline{\RV{{\beta}}}_{n}^{(j)} \triangleq [\overline{\RV{{\beta}}}_{1,n}^{(j)\mathrm{T}} \ist\cdots\ist \overline{\RV{{\beta}}}_{\rv{M}_{n}^{(j)},n}^{{(j)}\mathrm{T}} ]^{\mathrm{T}} $, respectively.

\subsection{Legacy PSFVs, Legacy PRs and State Transition}
\acp{pmva} that have been detected either at a previous time $n'<n$ or at time $n$ but at a previous \ac{pa} $j'<j$ are referred to as legacy \acp{pmva} with states $\underline{\RV{y}}_{s,n}^{(j)} \triangleq [{\underline{\RV{p}}_{s,\mathrm{\sfv}}^{(j)\mathrm{T}}} \iist \underline{\rv{r}}_{s,n}^{(j)}]^{\mathrm{T}} $. Accordingly, \acp{pva} that have been detected at a previous time $n'<n$ are referred to as legacy \acp{pva} with states $ \underline{\RV{{\beta}}}_{ss',n}^{(j)} \triangleq [ \underline{\rv{{u}}}_{ss',n}^{(j)} \iist \underline{\rv{r}}_{ss',n}^{(j)}]^{\mathrm{T}}$. Assume that measurements are incorporated sequentially across \acp{pa} $j \in \{1,\cdots, J\}$ at each time $n$. New \acp{pmva} become legacy \acp{pmva} after the next measurements either of the next \ac{pa} or at the next time are observed. New \acp{pva} of \ac{pa} $j$ become legacy \acp{pva} after the measurements of \ac{pa} $j$ at the next time are observed. The number of legacy \acp{pmva} is updated according to $ S_{n}^{(j)} = S_{n}^{(j-1)} + M_{n}^{(j-1)} $ for $j>1$ or $ S_{n}^{(1)} = S_{n-1}^{(J)} + M_{n-1}^{(J)} $. Accordingly, the number of \acp{pva} for each \ac{pa} is given as $ \vert \mathcal{D}_{n}^{(j)} \vert = S_{n}^{(j)} +S_{n}^{(j)} (S_{n}^{(j)} -1) $ with $ \mathcal{D}_{n}^{(j)} \triangleq \{(s\rmv,\rmv s') \in \mathcal{S}_{n}^{(j)}\rmv\rmv\times\rmv\rmv\mathcal{S}_{n}^{(j)} \} =  \mathcal{D}_{\mathrm{S},n}^{(j)} \cup \mathcal{D}_{\mathrm{D},n}^{(j)} $, and $\tilde{\mathcal{D}}_{n}^{(j)} = (0,\rmv 0)\rmv \cup \rmv \mathcal{D}_{n}^{(j)}$. Up to \ac{pa} $j$ at time $n$, the vector of all legacy \acp{pmva} is given as $ \underline{\RV{y}}_{n}^{(j)} \triangleq [\underline{\RV{y}}_{n}^{{(j-1)}\mathrm{T}} \iist \overline{\RV{y}}_{n}^{{(j-1)}\mathrm{T}} ]^{\mathrm{T}} $, and the vector of all \acp{pmva} is given as $ \RV{y}_{n}^{(j)} \triangleq [\underline{\RV{y}}_{n}^{{(j)}\mathrm{T}} \iist \overline{\RV{y}}_{n}^{{(j)}\mathrm{T}} ]^{\mathrm{T}} $. 
After the measurements of all $J$ \acp{pa} have been incorporated at time $n$, the total number of \acp{pmva} is $ S_{n} = S_{n-1} + \sum_{j=1}^{J} M_{n}^{(j)} = S_{n}^{(J)} + M_{n}^{(J)}$, the state of all \ac{pmva} states at time $n$ is given by $ \RV{y}_{n} \triangleq [\underline{\RV{y}}_{n}^{{(J)}\mathrm{T}} \iist \overline{\RV{y}}_{n}^{{(J)}\mathrm{T}} ]^{\mathrm{T}} $ which is further stacked into the vector for all times up to $n$ given by $ \RV{y}_{1:n} \triangleq [\RV{y}_{1}^{\mathrm{T}} \ist\cdots\ist \RV{y}_{n}^{\mathrm{T}} ]^{\mathrm{T}} $.
\vspace*{-0.5mm}

The agent state, legacy \acp{pmva} states and the legacy \ac{pva} states are assumed to evolve independently across time $n$ according to state-transition \acp{pdf} $f(\V{x}_{n}|\V{x}_{n-1})$, $f(\underline{\V{y}}_{s,n}|\V{y}_{s,n-1})$ and $f(\underline{\V{\beta}}_{ss',n}^{(j)} | \V{\beta}_{ss',n-1}^{(j)}) $. In line with \cite{Florian_Proceeding2018,Erik_SLAM_TWC2019, LeiVenTeaMey:TSP2023}, the state-transition \ac{pdf} $ f(\underline{\V{y}}_{s,n}|\V{y}_{s,n-1}) = f(\underline{\V{p}}_{s,\mathrm{\sfv}}, \underline{r}_{s,n}|\V{p}_{s,\mathrm{\sfv}}, r_{s,n-1}) $ at \ac{pa} $j=1$ is defined as
\begin{align}
	\hspace*{-3mm}f(\underline{\V{p}}_{s,\mathrm{\sfv}}, \underline{r}_{s,n}|\V{p}_{s,\mathrm{\sfv}}, 1) =
	\begin{cases}
		(1-p_{\mathrm{s}})f_{\mathrm{D}}(\underline{\V{p}}_{s,\mathrm{\sfv}}), 	&\underline{r}_{s,n}= 0\\
		p_{\mathrm{s}} f(\underline{\V{p}}_{s,\mathrm{\sfv}}|\V{p}_{s,\mathrm{\sfv}}), 											&\underline{r}_{s,n}= 1  
	\end{cases}
	\label{eq:STpdf_MVA1} 
\end{align}
for $r_{s,n-1} = 1$. For nonexistent \acp{pmva} at the previous time, i.e., $r_{s,n-1} = 0$, it is defined as
\begin{align}
 f(\underline{\V{p}}_{s,\mathrm{\sfv}}, \underline{r}_{s,n}|\V{p}_{s,\mathrm{\sfv}}, 0) =
	\begin{cases}
		f_{\mathrm{D}}(\underline{\V{p}}_{s,\mathrm{\sfv}}), 	&\underline{r}_{s,n}= 0\\
		0, 											&\underline{r}_{s,n}= 1 \, .
	\end{cases}
	\label{eq:STpdf_MVA2} 
\end{align}

Considering the sequential incorporation of \acp{pmva} states across \acp{pa} at each time $n$, we also define the state-transition \ac{pdf} $f(\underline{\V{y}}_{s,n}^{(j)}|\V{y}_{s,n}^{(j-1)}) = f(\underline{\V{p}}_{s,\mathrm{\sfv}}^{(j)}, \underline{r}_{s,n}^{(j)}|\V{p}_{s,\mathrm{\sfv}}^{(j-1)}, r_{s,n}^{(j-1)})$ for $j>1$, reads
\begin{align}
	& \hspace*{-3.3mm}  f(\underline{\V{p}}_{s,\mathrm{\sfv}}^{(j)}, \underline{r}_{s,n}^{(j)}|\V{p}_{s,\mathrm{\sfv}}^{(j-1)}, 1) =
	\begin{cases}
		f_{\mathrm{D}}(\underline{\V{p}}_{s,\mathrm{\sfv}}^{(j)}), 	& \hspace*{-0.7mm} \underline{r}_{s,n}^{(j)}= 0\\
		f(\underline{\V{p}}_{s,\mathrm{\sfv}}^{(j)}|\V{p}_{s,\mathrm{\sfv}}^{(j-1)}), 											& \hspace*{-0.7mm} \underline{r}_{s,n}^{(j)}= 1 \, ,
	\end{cases}
	\label{eq:STpdf_MVA3}  \vspace*{0mm}\\
	& \hspace*{-3.3mm} f(\underline{\V{p}}_{s,\mathrm{\sfv}}^{(j)}, \underline{r}_{s,n}^{(j)}|\V{p}_{s,\mathrm{\sfv}}^{(j-1)}, 0) =
	\begin{cases}
		f_{\mathrm{D}}(\underline{\V{p}}_{s,\mathrm{\sfv}}^{(j)}), 	&\underline{r}_{s,n}^{(j)}= 0\\
		0, 											&\underline{r}_{s,n}^{(j)}= 1 \, .
	\end{cases}
	\label{eq:STpdf_MVA4} 
\end{align}

The state-transition \ac{pdf} for \acp{pva} $f(\underline{\V{\beta}}_{ss',n}^{(j)} | \underline{\V{\beta}}_{ss',n-1}^{(j)}) = f(\underline{{u}}_{ss',n}^{(j)}, \underline{r}_{ss',n}^{(j)} | {u}_{ss',n-1}^{(j)}, r_{ss',n-1}^{(j)})$ is given by
\begin{align}
	& f(\underline{{u}}_{ss',n}^{(j)}, \underline{r}_{ss',n}^{(j)} | {u}_{ss',n-1}^{(j)}, r_{ss',n-1}^{(j)}=1) \nn \\[1mm]
	& \hspace*{14mm} =
	\begin{cases}
		(1-p_{\mathrm{s}})f_{\mathrm{D}}(\underline{{u}}_{ss',n}^{(j)}), &\underline{r}_{ss',n}^{(j)}= 0\\
		p_{\mathrm{s}} f(\underline{{u}}_{ss',n}^{(j)}|{u}_{ss',n-1}^{(j)}), 											&\underline{r}_{ss',n}^{(j)}= 1
	\end{cases}
	\label{eq:STpdf_VA1} 
\end{align}
for existing \ac{pva} at time $n-1$. For nonexistent \ac{pva} at the previous time, i.e., $r_{ss',n-1}^{(j)} = 0$, it is obtained as
\begin{align}
	\hspace*{-2mm} f(\underline{{u}}_{ss',n}^{(j)}, \underline{r}_{ss',n}^{(j)} | {u}_{ss',n-1}^{(j)}, 0)  =
	\begin{cases}
		f_{\mathrm{D}}(\underline{{u}}_{ss',n}^{(j)}), & \hspace*{-1.5mm} \underline{r}_{ss',n}^{(j)}= 0\\
		0, 											&\hspace*{-1.5mm} \underline{r}_{ss',n}^{(j)}= 1 \, .
	\end{cases}
	\label{eq:STpdf_VA2} 
\end{align}

The initial prior \acp{pdf} $f(\V{x}_{0})$, $f(\V{y}_{s,0})$ and $f(\V{\beta}_{ss',0}^{(j)})$ for $ s \in \mathcal{S}_{0} $, $ (s\rmv,\rmv s') \in \tilde{\mathcal{D}}_{0}^{(j)} $ and $j \in \{1,\dots,J\}$ at time $n=0$ are assumed to be known. 

\subsection{Data Association Uncertainty}
Estimation of multiple \ac{pmva} states is complicated by the \ac{da} uncertainty, i.e., it is unknown which measurement $\V{z}_{m,n}^{(j)}$ originated from which \ac{pa}, \ac{pmva}, or \ac{pmva} pair. Following \cite{LeiVenTeaMey:TSP2023}, the associations between measurements and legacy \acp{pmva} are described by the \emph{\ac{pmva}-oriented} association variables $ \RV{\underline{a}}_{n}^{(j)} \triangleq [\rv{\underline{a}}_{00,n}^{(j)} \ist \rv{\underline{a}}_{11,n}^{(j)} \ist \cdots \ist  \rv{\underline{a}}_{\rv{S}_{n}^{(j)}\rmv\rv{S}_{n}^{(j)},n}^{(j)}]^{\mathrm{T}} $ with entries $ \rv{\underline{a}}_{ss',n}^{(j)} \triangleq m \rmv\in\rmv \{1,\dots, \rv{M}_n^{(j)}\}$ if the legacy \ac{pmva} pair $ ss' $ generates measurement $ m $, or $ \rv{\underline{a}}_{ss',n}^{(j)} \triangleq 0 $ if the pair $ ss' $ does not generate any measurement. Following \cite{WilliamsLauTAE2014, Florian_Proceeding2018, Erik_SLAM_TWC2019}, the associations can be equivalently described by the \emph{measurement-oriented} association variables $ \RV{\overline{a}}_{n}^{(j)} \triangleq [\rv{\overline{a}}_{1,n}^{(j)} \ist \cdots \ist \rv{\overline{a}}_{\rv{M}_{n}^{(j)},n}^{(j)}]^{\mathrm{T}} $ with entries $ \rv{\overline{a}}_{m,n}^{(j)} \triangleq (s\rmv,\rmv s') \in \tilde{\mathcal{D}}_{n}^{(j)}$ if measurement $ m $ is generated by legacy \ac{pmva} pair $ ss' $, or $ \rv{\overline{a}}_{m,n}^{(j)}  \triangleq 0 $ if measurement $ m $ is not generated by any legacy \ac{pmva}. Note that the association variables are valid for all cases with $ (s\rmv,\rmv s') \in \tilde{\mathcal{D}}_{n}^{(j)} $. At any time $ n $, we follow the assumption that each \ac{pva} can generate at most one measurement, and each measurement can be generated by at most one \ac{pmva} pair, which is guaranteed by the ``redundant formulation'' of using $ \RV{\underline{a}}_{n}^{(j)} $ together with $ \RV{\overline{a}}_{n}^{(j)} $. This is also the key to make the algorithm scalable for large numbers of \acp{pmva} and measurements. The joint association vectors for all \acp{pa} and for all times up to $n$ are given by 
$\RV{\underline{a}}_{n} \triangleq [\RV{\underline{a}}_{n}^{(1)\mathrm{T}} \ist\cdots\ist \RV{\underline{a}}_{n}^{(J)\mathrm{T}} ]^{\mathrm{T}}$, $ \RV{\overline{a}}_{n} \triangleq [\RV{\overline{a}}_{n}^{(1)\mathrm{T}} \ist\cdots\ist \RV{\overline{a}}_{n}^{(J)\mathrm{T}} ]^{\mathrm{T}} $, $\RV{\underline{a}}_{1:n} \triangleq [\RV{\underline{a}}_{1}^{\mathrm{T}} \ist\cdots\ist \RV{\underline{a}}_{n}^{\mathrm{T}} ]^{\mathrm{T}}$ and $ \RV{\overline{a}}_{1:n} \triangleq [\RV{\overline{a}}_{1}^{\mathrm{T}} \ist\cdots\ist \RV{\overline{a}}_{n}^{\mathrm{T}} ]^{\mathrm{T}} $.

%% file: InputFiles/ProblemFormulation.tex
\begin{figure*}[!t]	
	\begin{align}
		& \hspace*{0mm} f(\V{x}_{0:n}, \V{y}_{0:n}, \V{\beta}_{0:n}, \V{\underline{a}}_{1:n}, \V{\overline{a}}_{1:n} | \V{z}_{1:n}) \nn \\
		& \hspace*{-2mm}\propto \underbrace{ \left( f(\V{x}_{0}) \prod_{s = 1}^{S_{0}} f(\V{y}_{s,0}) \prod_{j'' = 1}^{J} \prod_{(s\rmv,\rmv s') \in \tilde{\mathcal{D}}_{0}^{(j'')}} \rrmv \rrmv f(\V{\beta}_{ss',0}^{(j'')}) \right) }_{\text{initial prior \acp{pdf}}} 
		\underbrace{ \prod_{n' = 1}^{n} f(\V{x}_{n'}|\V{x}_{n'-1}) }_{\text{agent state transition}} \underbrace{ \left( \prod_{s' = 1}^{S_{n'-1}} f(\underline{\V{y}}_{s',n'}|\V{y}_{s',n'-1}) \right) \rrmv \left( \prod_{j' = 2}^{J}  \prod_{s' = 1}^{S_{n'}^{(j')}} f^{(j)}(\underline{\V{y}}^{(j')}_{s',n'}|\V{y}^{(j'-1)}_{s',n'}) \right) }_{\text{legacy \acp{pmva}' state transition}} \nn \\ & \hspace*{2mm}
		\times \underbrace{ \left( \prod_{j = 1}^{J} f(\underline{\V{\beta}}_{00,n'}^{(j)} | \V{\beta}_{00,n'-1}^{(j)}) \underline{q}_{\mathrm{P}}(\V{x}_{n'}, \underline{\V{\beta}}_{00,n'}^{(j)}, \underline{a}_{00,n'}^{(j)};\V{z}_{n'}^{(j)}) \prod_{m' = 1}^{M_{n'}^{(j)}} \psi(\underline{a}_{00,n'}^{(j)},\overline{a}_{m',n'}^{(j)}) \right) }_{\text{factors related to \acp{pa} and \ac{los} \acp{pva}}}  \nn \\ & \hspace*{2mm}
		\times 
		\hspace*{1mm} \underbrace{ \prod_{j = 1}^{J} \left( \prod_{s = 1}^{S_{n'}^{(j)}} f(\underline{\V{\beta}}_{ss,n'}^{(j)} | \V{\beta}_{ss,n'-1}^{(j)}) \underline{q}_{\mathrm{S}}(\V{x}_{n'}, \underline{\V{y}}_{s,n'}^{(j)},  \underline{\V{\beta}}_{ss,n'}^{(j)}, \underline{a}_{ss,n'}^{(j)};\V{z}_{n'}^{(j)}) \prod_{m' = 1}^{M_{n'}^{(j)}} \psi(\underline{a}_{ss,n'}^{(j)},\overline{a}_{m',n'}^{(j)}) \right) }_{\text{factors related to legacy \acp{pmva} and single-bounce \acp{pva}}} 
		\hspace*{2mm} 
		\nn \\ & \hspace*{2mm} 
		\times 
		\underbrace{ \left(  \prod_{s' =1,s'\neq s}^{S_{n'}^{(j)}} f(\underline{\V{\beta}}_{ss',n'}^{(j)} | \V{\beta}_{ss',n'-1}^{(j)}) \underline{q}_{\mathrm{D}}(\V{x}_{n'}, \underline{\V{y}}_{s,n'}^{(j)}, \underline{\V{y}}_{s',n'}^{(j)},  \underline{\V{\beta}}_{ss',n'}^{(j)}, \underline{a}_{ss',n'}^{(j)};\V{z}_{n'}^{(j)}) \prod_{m' = 1}^{M_{n'}^{(j)}} \psi(\underline{a}_{ss',n'}^{(j)},\overline{a}_{m',n'}^{(j)}) \right) }_{\text{factors related to legacy \acp{pmva} and double-bounce \acp{pva}}} 
		\nn \\ & \hspace*{2mm} 
		\times \underbrace{ \left( \prod_{m = 1}^{M_{n'}^{(j)}} \overline{q}_{\mathrm{N}}(\V{x}_{n'}, \overline{\V{y}}_{m,n'}^{(j)}, \overline{\V{\beta}}_{m,n'}^{(j)}, \overline{a}_{m,n'}^{(j)};\V{z}_{n'}^{(j)}) \right) }_{\text{prior \acp{pdf} and factors related to new \acp{pmva} and \acp{pva} }} 
		\label{eq:jointPDF}
	\end{align}
	\hrulefill
	\vspace*{-4mm}
\end{figure*}

Using all the measurements $ \RV{z}_{1:n} $ of all \acp{pa} and all times up to $n$, we aim to sequentially detect the existences of \acp{pva} and \acp{pmva}, and estimate their states $\RV{{\beta}}^{(j)}_{ss',n}$, $\RV{y}_{s,n}$ and the agent state $\RV{x}_{n}$. This relies on the marginal posterior existence probabilities $f(r_{ss',n}^{(j)}=1 | \V{z}_{1:n}) =  \int f( u_{ss',n}^{(j)}, r_{ss',n}^{(j)}=1 | \V{z}_{1:n}^{(\sim j)}) \mathrm{d} u_{ss',n}^{(j)} $ with $\V{z}_{1:n}^{(\sim j)} \triangleq [\V{z}_{1:n-1}^{\mathrm{T}} \quad [\RV{z}_{n}^{{(1)}\mathrm{T}} \ist \cdots \ist \RV{z}_{n}^{{(j)}\mathrm{T}}] ]^{\mathrm{T}}$ and $ f( r_{s,n}=1 | \V{z}_{1:n}) = \int f( \V{p}_{s,\mathrm{\sfv}}, r_{s,n}=1 |\V{z}_{1:n}) \mathrm{d} \V{p}_{s,\mathrm{\sfv}} $, and the marginal posterior \acp{pdf} $ f( u_{ss',n}^{(j)} |r_{ss',n}^{(j)}=1, \V{z}_{1:n}^{(\sim j)}) = f( u_{ss',n}^{(j)}, r_{ss',n}^{(j)}=1 | \V{z}_{1:n}^{(\sim j)})/f(r_{ss',n}^{(j)}=1 | \V{z}_{1:n}) $ and $f( \V{p}_{s,\mathrm{\sfv}} |r_{s,n}=1,\V{z}_{1:n}) = f( \V{p}_{s,\mathrm{\sfv}}, r_{s,n}=1 |\V{z}_{1:n})/ f( r_{s,n}=1 | \V{z}_{1:n}) $.

\subsection{Existence Detection and State Estimation}
\begin{figure*}[!t]
	\centering
	\hspace*{-0mm}\subfloat[\label{subfig:flowchart}]
	{\hspace*{-2mm}\includegraphics[scale=1.02]{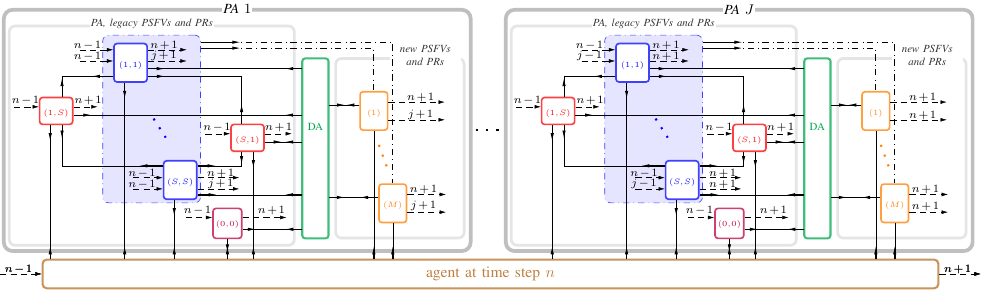}}\\[-4mm]
	\hspace*{5mm}\subfloat[\label{subfig:FGPA}]
	{\hspace*{-5mm}\includegraphics[scale=0.92]{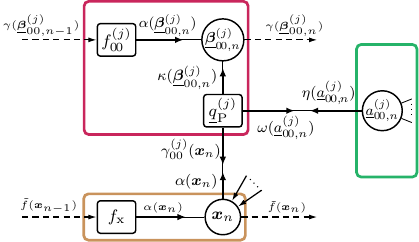}}
	\hspace*{20.5mm}\subfloat[\label{subfig:FGMVA1}]
	{\hspace*{-6.5mm}\includegraphics[scale=0.92]{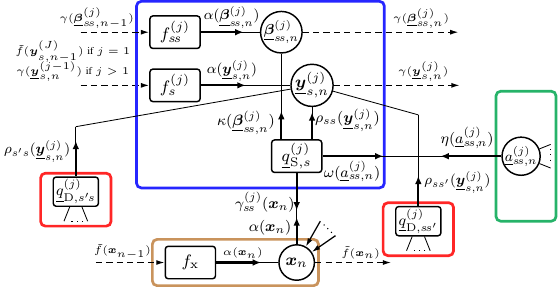}}\\[-6mm]
	\hspace*{-20.5mm}\subfloat[\label{subfig:FGMVA2}]
	{\hspace*{4.5mm}\includegraphics[scale=0.92]{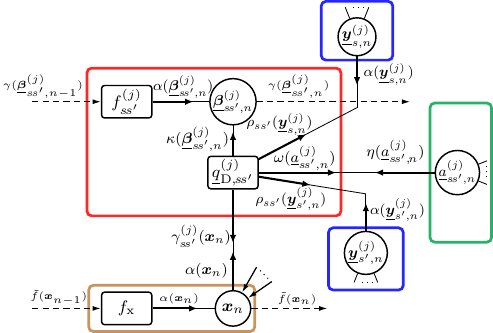}}
	\hspace*{26mm}\subfloat[\label{subfig:FGMVAnew}]
	{\hspace*{-14mm}\includegraphics[scale=0.92]{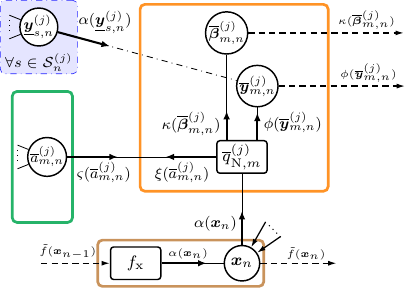}}\\[-1mm]
	\caption{Factor graph representation of the joint posterior \ac{pdf} in \eqref{eq:jointPDF}. The black square and circle symbols denote factor nodes and variable nodes, respectively, while the colored boxes represent subgraphs formed by subsets of nodes and edges. The block diagram in (a) illustrates message propagation between the subgraphs, as well as iterations over time and \acp{pa}. Detailed views of the variable nodes, factor nodes, and messages related to (b) \acp{pa} and \ac{los} \acp{pva}, (c) legacy \acp{pmva} and single-bounce \acp{pva}, (d) double-bounce \acp{pva}, and (e) new \acp{pmva} and new \acp{pva} are shown in the corresponding zoomed-in plots. The following shorthand notation is used: $S \triangleq S_{n}^{(j)}$, $M \triangleq M_{n}^{(j)}$, $f_{\mathrm{x}} \triangleq f(\V{x}_{n}|\V{x}_{n-1})$, $f_{s}^{(j)} \triangleq f(\underline{\V{y}}_{s,n}^{(j)}|\V{y}_{s,n-1}^{(j)})$, $f_{s\rmv s'}^{(j)} \triangleq f(\underline{\V{\beta}}_{ss',n}^{(j)} | \V{\beta}_{ss',n-1}^{(j)})$ with $(s\rmv,\rmv s') \in \tilde{\mathcal{D}}_{n}^{(j)}$, and $ \underline{q}_{\mathrm{P}}^{(j)} $, $ \underline{q}_{\mathrm{S},s}^{(j)} $, $ \underline{q}_{\mathrm{D},\rmv s\rmv s'}^{(j)} $, and $ \overline{q}_{\mathrm{N},m}^{(j)} $ denote the pseudo \acp{lhf} of the \ac{los} paths \eqref{eq:LHFPA}, single-bounce paths \eqref{eq:LHFSpath}, double-bounce paths \eqref{eq:LHFDpath}, and new paths \eqref{eq:LHFNew}, respectively. For a detailed factor graph representation of the \ac{da} component, the reader is referred to \cite{LeiVenTeaMey:TSP2023}. In (e), new \ac{pmva} states are connected to the legacy \ac{pmva} states, reflecting the use of predicted legacy \ac{pmva} states for constructing the proposal \ac{pdf} of new \acp{pmva} supported by double-bounce paths, as detailed in Section~\ref{subsec:ImpAsp}\vspace{2mm}.}
	\label{fig:FactorGraphSum}
	\vspace*{0mm}
\end{figure*}

The agent state is estimated by means of the \ac{mmse} estimator \cite{Kay_EstimationTheory}, i.e., 
\begin{align}	
	\hat{\V{x}}_{n} \triangleq \int \V{x}_{n} f(\V{x}_{n}|\V{z}_{1:n})\mathrm{d} \V{x}_{n}.
	\label{eq:MMSE_x} 
	\vspace{0mm}
\end{align}
A \ac{pva} and a \ac{pmva} are declared to exist (i.e., detected) if $ f( r_{ss',n}^{(j)}=1 | \V{z}_{1:n}) > p_{\mathrm{\pr}}$ and $ f( r_{s,n}=1 | \V{z}_{1:n}) > p_{\mathrm{\sfv}}$, where $p_{\mathrm{\pr}}$ and $p_{\mathrm{\sfv}}$ are the existence probability thresholds. For existing \acp{pmva} and \acp{pva}, their positions and normalized amplitudes are again calculated by the \ac{mmse}, yielding
\begin{align}	
	\hat{\V{p}}_{s,\mathrm{\sfv}} &\triangleq \int \V{p}_{s,\mathrm{\sfv}}  f( \V{p}_{s,\mathrm{\sfv}} |r_{s,n}=1,\V{z}_{1:n}) \mathrm{d} \V{p}_{s,\mathrm{\sfv}},
	\label{eq:MMSE_PMVA} 	\\
	\hat{u}_{ss',n}^{(j)} &\triangleq \int u_{ss',n}^{(j)} f( u_{ss',n}^{(j)} |r_{ss',n}^{(j)}=1, \V{z}_{1:n}^{(\sim j)}) \mathrm{d} u_{ss',n}^{(j)}.
	\label{eq:MMSE_PVA} 
	\vspace{0mm}
\end{align}
The posterior \acp{pdf} $f(\V{x}_{n}|\V{z}_{1:n})$,  $f( \V{p}_{s,\mathrm{\sfv}}, r_{s,n}=1|\V{z}_{1:n})$ and $ f( u_{ss',n}^{(j)},r_{ss',n}^{(j)}=1| \V{z}_{1:n}^{(\sim j)})$ are marginal \acp{pdf} of the joint posterior \ac{pdf} $f(\V{x}_{0:n}, \V{y}_{0:n}, \V{\beta}_{0:n}, \V{\underline{a}}_{1:n}, \V{\overline{a}}_{1:n} | \V{z}_{1:n})$. Since the direct marginalization of the joint posterior \ac{pdf} is computationally infeasible, we perform message passing using the \ac{spa} rules on the factor graph in Fig.\,\ref{fig:FactorGraphSum} representing the joint posterior \ac{pdf} in \eqref{eq:jointPDF}, which efficiently calculates the beliefs $\tilde{f}(\V{x}_{n})$, $\tilde{f}_{s}(\V{y}_{s,n})$ and $\tilde{f}(\V{{\beta}}^{(j)}_{ss',n})$ approximating these marginal \acp{pdf}\vspace{-2mm}.

\subsection{Joint Posterior \ac{pdf} and Factor Graph}
\label{sec:jointPDFandFG}
Using common assumptions \cite{BarShalom_AlgorithmHandbook, Florian_Proceeding2018, Erik_SLAM_TWC2019}, the joint posterior \ac{pdf} $f(\V{x}_{0:n}, \V{y}_{0:n}, \V{\beta}_{0:n}, \V{\underline{a}}_{1:n}, \V{\overline{a}}_{1:n} | \V{z}_{1:n})$ of $ \RV{x}_{0:n} $, $ \RV{y}_{0:n} $, $ \RV{{\beta}}_{0:n} $, $ \RV{\underline{a}}_{1:n} $ and $ \RV{\overline{a}}_{1:n} $ conditioned on the observed (thus fixed) measurements $ \V{z}_{1:n} $ is factorized as \eqref{eq:jointPDF}, where the functions $\underline{q}_{\mathrm{P}}(\V{x}_{n}, \underline{\V{\beta}}_{00,n}^{(j)}, \underline{a}_{00,n}^{(j)};\V{z}_{n}^{(j)})$, $\underline{q}_{\mathrm{S}}(\V{x}_{n}, \underline{\V{y}}_{s,n}^{(j)}, \underline{\V{\beta}}_{ss,n}^{(j)}, \underline{a}_{ss,n}^{(j)};\V{z}_{n}^{(j)}) $, $\underline{q}_{\mathrm{D}}(\V{x}_{n}, \underline{\V{y}}_{s,n}^{(j)}, \underline{\V{y}}_{s',n}^{(j)}, \underline{\V{\beta}}_{ss',n}^{(j)}, \underline{a}_{ss',n}^{(j)}; \V{z}_{n}^{(j)}) $ and $\overline{q}_{\mathrm{N}}(\V{x}_{n}, \overline{\V{y}}_{m,n}^{(j)}, \\ \overline{\V{\beta}}_{m,n}^{(j)}, \overline{a}_{m,n}^{(j)};\V{z}_{n}^{(j)})$ will be discussed next.

\subsubsection*{For the \ac{los} \ac{pva} related to \ac{pa} $j$} the pseudo \ac{lhf} $\underline{q}_{\mathrm{P}}(\V{x}_{n}, \underline{\V{\beta}}_{00,n}^{(j)}, \underline{a}_{00,n}^{(j)};\V{z}_{n}^{(j)}) = \underline{q}_{\mathrm{P}}(\V{x}_{n}, \underline{u}_{00,n}^{(j)}, \underline{r}_{00,n}^{(j)}, \underline{a}_{00,n}^{(j)};\V{z}_{n}^{(j)})$ is given by
\begin{align}
	& \hspace*{0mm}\underline{q}_{\mathrm{P}}(\V{x}_{n}, \underline{u}_{00,n}^{(j)}, \underline{r}_{00,n}^{(j)}=1, \underline{a}_{00,n}^{(j)};\V{z}_{n}^{(j)}) \nn \\[1mm]
	& \hspace*{0mm} \triangleq
	\begin{cases}
		\dfrac{ f_{\mathrm{P}}(\V{z}_{m,n}^{(j)}|\V{p}_{n}, \underline{u}_{00,n}^{(j)})  p_{\mathrm{d}}(\underline{u}_{00,n}^{(j)}) } {\mu_{\mathrm{fa}} f_{\mathrm{fa}}(\V{z}_{m,n}^{(j)})}, 										& \underline{a}_{00,n}^{(j)} \in \mathcal{M}_{n}^{(j)} \\[4mm]
		1 - p_{\mathrm{d}}(\underline{u}_{00,n}^{(j)}),										& \underline{a}_{00,n}^{(j)} = 0.
	\end{cases}
	\label{eq:LHFPA} 
\end{align}
and $ \underline{q}_{\mathrm{P}}(\V{x}_{n}, \underline{u}_{00,n}^{(j)}, \underline{r}_{00,n}^{(j)}=0, \underline{a}_{00,n}^{(j)};\V{z}_{n}^{(j)}) \triangleq \delta(\underline{a}_{00,n}^{(j)}) $.

\subsubsection*{For single-bounce \acp{pva}} related to \ac{pmva} $(s\rmv,\rmv s) \in \mathcal{D}_{\mathrm{S},n}^{(j)}$, the pseudo \ac{lhf} $\underline{q}_{\mathrm{S}}(\V{x}_{n}, \underline{\V{y}}_{s,n}^{(j)}, \underline{\V{\beta}}_{ss,n}^{(j)}, \underline{a}_{ss,n}^{(j)};\V{z}_{n}^{(j)}) = \underline{q}_{\mathrm{S}}(\V{x}_{n}, \underline{\V{p}}_{s,\mathrm{\sfv}}^{(j)}, \underline{r}_{s,n}^{(j)}, \underline{u}_{ss,n}^{(j)},\underline{r}_{ss,n}^{(j)}, \underline{a}_{ss,n}^{(j)};\V{z}_{n}^{(j)})$ is given by 
\begin{align}
	& \hspace*{0mm} \underline{q}_{\mathrm{S}}(\V{x}_{n}, \underline{\V{p}}_{s,\mathrm{\sfv}}^{(j)}, \underline{r}_{s,n}^{(j)} = 1, \underline{u}_{ss,n}^{(j)},\underline{r}_{ss,n}^{(j)}= 1, \underline{a}_{ss,n}^{(j)};\V{z}_{n}^{(j)}) \nonumber \\[2mm]
	& \hspace*{-5mm} \triangleq 
	\begin{cases}
		\dfrac{ f_{\mathrm{S}}(\V{z}_{m,n}^{(j)}|\V{p}_{n}, \underline{\V{p}}_{s,\mathrm{\sfv}}^{(j)},  \underline{u}_{ss,n}^{(j)})  p_{\mathrm{d}}(\underline{u}_{ss,n}^{(j)}) } {\mu_{\mathrm{fa}} f_{\mathrm{fa}}(\V{z}_{m,n}^{(j)})}, 								& \hspace*{-2mm} \underline{a}_{ss,n}^{(j)} \in \mathcal{M}_{n}^{(j)} \\[4mm]
		1 - p_{\mathrm{d}}(\underline{u}_{ss,n}^{(j)}),						& \hspace*{-2mm} \underline{a}_{ss,n}^{(j)} = 0
	\end{cases}
	\hspace*{-2mm}\label{eq:LHFSpath}
\end{align}
and $\underline{q}_{\mathrm{S}}(\V{x}_{n}, \underline{\V{p}}_{s,\mathrm{\sfv}}^{(j)}, \underline{r}_{s,n}^{(j)} = 0, \underline{u}_{ss,n}^{(j)},\underline{r}_{ss,n}^{(j)}= 1, \underline{a}_{ss,n}^{(j)}; \V{z}_{n}^{(j)}) = 0$ for the invalid case, i.e., \ac{pmva} $s$ does not exist but \ac{pva} $(s\rmv,\rmv s)$ exists, otherwise $\underline{q}_{\mathrm{S}}(\cdots)  \triangleq \delta(\underline{a}_{ss,n}^{(j)})$. 
\vspace*{-2mm}

\subsubsection*{For double-bounce \acp{pva}} related to \ac{pmva} pair $(s\rmv,\rmv s') \in \mathcal{D}_{\mathrm{D},n}^{(j)}$, the pseudo \ac{lhf} $\underline{q}_{\mathrm{D}}(\V{x}_{n}, \underline{\V{y}}_{s,n}^{(j)},\underline{\V{y}}_{s',n}^{(j)}, \underline{\V{{\beta}}}_{ss',n}^{(j)}, \underline{a}_{ss',n}^{(j)};  \V{z}_{n}^{(j)}) \\ =  \underline{q}_{\mathrm{D}}(\V{x}_{n}, \underline{\V{p}}_{s,\mathrm{\sfv}}^{(j)},  \underline{r}_{s,n}^{(j)},  \underline{\V{p}}_{s',\mathrm{\sfv}}^{(j)}, \underline{r}_{s',n}^{(j)}, \underline{u}_{ss',n}^{(j)}, \underline{r}_{ss',n}^{(j)}, \underline{a}_{ss',n}^{(j)};\V{z}_{n}^{(j)}) $ is given by
\vspace*{-2mm}
\begin{align}
	& \hspace*{0mm} \underline{q}_{\mathrm{D}}(\V{x}_{n}, \underline{\V{p}}_{s,\mathrm{\sfv}}^{(j)}, 1, \underline{\V{p}}_{s',\mathrm{\sfv}}^{(j)}, 1, \underline{u}_{ss',n}^{(j)}, 1, \underline{a}_{ss',n}^{(j)};\V{z}_{n}^{(j)}) \nonumber \\[1mm]
	& \hspace*{0mm} \triangleq 
	\begin{cases}
		 f_{\mathrm{D}}(\V{z}_{m,n}^{(j)}|\V{p}_{n}, \underline{\V{p}}_{s,\mathrm{\sfv}}^{(j)}, \underline{\V{p}}_{s',\mathrm{\sfv}}^{(j)},  \underline{u}_{ss',n}^{(j)})  \\ 
		\hspace*{0mm} \hspace{0.3em} \times \hspace{0.3em} 
		 \dfrac{p_{\mathrm{d}}(\underline{u}_{ss',n}^{(j)})} {\mu_{\mathrm{fa}}  f_{\mathrm{fa}}(\V{z}_{m,n}^{(j)})}, 										& \hspace*{-6mm} \underline{a}_{ss',n}^{(j)} \in \mathcal{M}_{n}^{(j)} \\[4mm]
		1 - p_{\mathrm{d}}(\underline{u}_{ss',n}^{(j)}),										& \hspace*{-6mm} \underline{a}_{ss',n}^{(j)} = 0
	\end{cases}
	\label{eq:LHFDpath} 
\end{align}
and $ \underline{q}_{\mathrm{D}}(\cdots) = 0$ for $( \underline{r}_{s,n}^{(j)}, \underline{r}_{s',n}^{(j)}, \underline{r}_{ss',n}^{(j)}) \in \{ (0,1,1),(1,0,1), \\ (0,0,1) \}$ representing invalid cases that path $(s\rmv,\rmv s')$ exists but one or both the associated surfaces are nonexistent. For other cases, $ \underline{q}_{\mathrm{D}}(\cdots) \triangleq \delta(\underline{a}_{ss',n}^{(j)})$. 

\subsubsection*{For new \acp{pva}} related to new \acp{pmva}, the pseudo \ac{lhf} $ \overline{q}_{\mathrm{N}}( \V{x}_{n}, \overline{\V{y}}_{m,n}^{(j)}, \overline{\V{\beta}}_{m,n}^{(j)}, \overline{a}_{m,n}^{(j)};\V{z}_{n}^{(j)}) = \overline{q}_{\mathrm{N}}(\V{x}_{n}, \overline{\V{p}}_{m,\mathrm{\sfv}}^{(j)},  \overline{r}_{m,n}^{(j)}, \\ \overline{u}_{mm,n}^{(j)}, \overline{r}_{mm,n}^{(j)}, \overline{a}_{m,n}^{(j)};\V{z}_{n}^{(j)}) $ is given by
\begin{align}
	& \hspace*{0mm} \overline{q}_{\mathrm{N}}(\V{x}_{n}, \overline{\V{p}}_{m,\mathrm{\sfv}}^{(j)}, \overline{r}_{m,n}^{(j)}=1, \overline{u}_{mm,n}^{(j)}, \overline{r}_{mm,n}^{(j)}=1, \overline{a}_{m,n}^{(j)};\V{z}_{n}^{(j)}) \nonumber \\[1mm]
	& \hspace*{-6.5mm} \triangleq 
	\begin{cases}
		0,										&\hspace*{0mm} \overline{a}_{m,n}^{(j)} \in \tilde{\mathcal{D}}_{n}^{(j)} \\[2mm]
		f_{\mathrm{N}}(\V{z}_{m,n}^{(j)}|\V{p}_{n}, \overline{\V{p}}_{m,\mathrm{\sfv}}^{(j)}, \overline{u}_{mm,n}^{(j)})   \\
		\hspace*{0mm} \hspace{0.3em} \times \hspace{0.3em} 
		\dfrac{ \mu_{\mathrm{n}} f_{\mathrm{n}}(\overline{\V{p}}_{m,\mathrm{\sfv}}^{(j)}, \overline{u}_{mm,n}^{(j)} | \V{x}_{n})} {\mu_{\mathrm{fa}} f_{\mathrm{fa}}(\V{z}_{m,n}^{(j)}) } 
		,  										&\hspace*{0mm} \overline{a}_{m,n}^{(j)} = 0
	\end{cases}
	\label{eq:LHFNew}
\end{align}
and $ \overline{q}_{\mathrm{N}}(\V{x}_{n},\rmv\overline{\V{p}}_{m,\mathrm{\sfv}}^{(j)}, \overline{r}_{m,n}^{(j)}=0,\rmv\overline{u}_{mm,n}^{(j)},\rmv\overline{r}_{mm,n}^{(j)}=0, \overline{a}_{m,n}^{(j)};\rmv\V{z}_{n}^{(j)}) \\ \triangleq f_{\mathrm{D}}(\overline{\V{p}}_{m,\mathrm{\sfv}}^{(j)}) f_{\mathrm{D}}(\overline{u}_{mm,n}^{(j)}) $. For other cases, $\overline{q}_{\mathrm{N}}(\cdots) = 0$. Details about the joint prior \ac{pdf} $f_{\mathrm{n}}(\overline{\V{p}}_{m,\mathrm{\sfv}}^{(j)}, \overline{u}_{mm,n}^{(j)} | \V{x}_{n})$ and the \ac{lhf} $f_{\mathrm{N}}(\V{z}_{m,n}^{(j)}|\V{p}_{n}, \overline{\V{p}}_{m,\mathrm{\sfv}}^{(j)}, \overline{u}_{mm,n}^{(j)}) \triangleq f_{\forall s}(\V{z}_{m,n}^{(j)}|\V{p}_{n}, \overline{\V{p}}_{m,\mathrm{\sfv}}^{(j)}, \\ \underline{\V{p}}_{s,\mathrm{\sfv}}^{(j)}, \overline{u}_{mm,n}^{(j)})$ are provided in Section~\ref{subsec:ImpAsp}.

The joint posterior \ac{pdf} in \eqref{eq:jointPDF} and its factor graph representation in Fig.~\ref{fig:FactorGraphSum}, the binary check function $\psi(\underline{a}_{ss',n}^{(j)},\overline{a}_{m,n}^{(j)})$ are in parts in line with \cite{Florian_Proceeding2018, Erik_SLAM_TWC2019, XuhongTWC2022, LeiVenTeaMey:TSP2023}. Details of the derivations of the system models, including \eqref{eq:jointPDF} and the pseudo \acp{lhf} \eqref{eq:LHFPA}-\eqref{eq:LHFNew} are provided in the supplementary material \cite[Section~I]{MVASLAM_TWC2025SupportingDoc}.

%% file: InputFiles/SPA.tex
Due to the loops inside the factor graph, we specify the following orders for message computation: (i) messages are sent only forward in time from $n-1$ to $n$ and serially from \ac{pa} $j-1$ to $j$; (ii) iterative message passing is only performed for probabilistic \ac{da}, i.e., message passing iteration is performed only once in the loops connecting different \acp{pmva}; (iii) messages are only sent from the agent state variable node to the new \ac{pmva} state variable node, not vice versa. Combining the specified orders with the generic \ac{spa} rules, the messages and beliefs involved in the factor graph in Fig.~\ref{fig:FactorGraphSum} are introduced for each time step $n$ and each \ac{pa} $j$ as follows. For brevity, we provide only the general steps of the \ac{spa} algorithm in the following. The derivations of each message are presented in detail in the supplementary material \cite[Section~II]{MVASLAM_TWC2025SupportingDoc}.

\subsection{Prediction} 
\subsubsection{Transition from Time $n-1$ to $n$}

First, a prediction step from time $n-1$ to $n$ is performed for the agent state, all legacy \ac{pmva} states and \ac{pva} states, leading to the predicted message for the agent state $\alpha(\V{x}_{n})$, the predicted messages $\alpha_{s}(\underline{\V{p}}_{s,\mathrm{\sfv}}, \underline{r}_{s,n})$ for all the legacy \acp{pmva} $s \in \mathcal{S}_{n-1}$ and the predicted messages $\alpha_{ss'}(\underline{\V{\beta}}_{ss',n}^{(j)}) = \alpha_{ss'}(\underline{u}_{ss',n}^{(j)}, \underline{r}_{ss',n}^{(j)})$ for legacy \acp{pva}.

\subsubsection{Transition from \ac{pa} $j-1$ to $j$}
For $j>1$ at time $n$, the predicted messages $\alpha_{s}(\underline{\V{p}}_{s,\mathrm{\sfv}}^{(j)}, \underline{r}_{s,n}^{(j)})$ for former legacy \acp{pmva} $s \in \mathcal{S}_{n}^{(j-1)}$ and the predicted messages $\alpha_{S_{n}^{(j-1)}+m}(\underline{\V{p}}_{S_{n}^{(j-1)}+m,\mathrm{\sfv}}^{(j)}, \underline{r}_{S_{n}^{(j-1)}+m,n}^{(j)})$ for former new \acp{pmva} $m \in \mathcal{M}_{n}^{(j-1)}$ are as in \cite{LeiVenTeaMey:TSP2023}.

\subsection{Sequential \ac{pa} Update}
\subsubsection{Measurement Evaluation for \ac{los} \acp{pva}}
The messages $\omega(\underline{a}_{00,n}^{(j)})$ propagating from the factor node $\underline{q}_{\mathrm{P}}(\V{x}_{n}, \underline{\V{\beta}}_{00,n}^{(j)}, \underline{a}_{00,n}^{(j)}; \V{z}_{n}^{(j)})$ to the feature-oriented association variable nodes $\underline{a}_{00,n}^{(j)}$ are calculated.
\subsubsection{Measurement Evaluation for Legacy \acp{pmva}}
For $s=s'$, the messages propagating from the factor node $\underline{q}_{\mathrm{S}}(\V{x}_{n}, \underline{\V{y}}_{s,n}^{(j)}, \underline{\V{\beta}}_{ss,n}^{(j)}, \underline{a}_{ss,n}^{(j)};\V{z}_{n}^{(j)}) $ to the feature-oriented association variable nodes $\underline{a}_{ss,n}^{(j)}$ are given as $\omega(\underline{a}_{ss,n}^{(j)})$. For $s\neq s'$, the messages propagating from the factor node $\underline{q}_{\mathrm{D}}(\V{x}_{n}, \underline{\V{y}}_{s,n}^{(j)},\underline{\V{y}}_{s',n}^{(j)}, \underline{\V{\beta}}_{ss',n}^{(j)}, \underline{a}_{ss',n}^{(j)};\V{z}_{n}^{(j)}) $ to the feature-oriented association variable nodes $\underline{a}_{ss',n}^{(j)}$ are given as $\omega(\underline{a}_{ss',n}^{(j)})$.

\subsubsection{Measurement Evaluation for New \acp{pmva}}
For PA $j$, the messages $\xi(\overline{a}_{m,n}^{(j)})$ sent from $ \overline{q}_{\mathrm{N}}( \V{x}_{n}, \overline{\V{y}}_{m,n}^{(j)}, \overline{\V{\beta}}_{m,n}^{(j)}, \overline{a}_{m,n}^{(j)};\V{z}_{n}^{(j)}) $ to the measurement-oriented association variable nodes $\overline{a}_{m,n}^{(j)}$ are calculated.

\subsubsection{Iterative Data Association}
\label{subsubsection:DA}
With the messages $\omega(\underline{a}_{ss',n}^{(j)}) $ and $ \xi(\overline{a}_{m,n}^{(j)}) $, the probabilistic \ac{da} messages $ \eta(\underline{a}_{ss',n}^{(j)}) $ and $\varsigma(\overline{a}_{m,n}^{(j)}) $ are obtained using loopy \ac{bp} according to \cite{Florian_Proceeding2018, LeiVenTeaMey:TSP2023}.

\subsubsection{Measurement Update for The Agent}
For the update of agent state $\V{x}_{n} $, only the messages from legacy \acp{pmva} are used. The messages sent from the factor node $\underline{q}_{\mathrm{P}}(\V{x}_{n}, \underline{\V{\beta}}_{00,n}^{(j)}, \underline{a}_{00,n}^{(j)};\V{z}_{n}^{(j)})$ to the agent state variable node $\V{x}_{n} $ are given as $\gamma^{(j)}_{00}(\V{x}_{n})$. For $s=s'$, the messages $\gamma^{(j)}_{ss}(\V{x}_{n})$ propagating from the factor node $\underline{q}_{\mathrm{S}}(\V{x}_{n}, \underline{\V{y}}_{s,n}^{(j)}, \underline{\V{\beta}}_{ss,n}^{(j)}, \underline{a}_{ss,n}^{(j)};\V{z}_{n}^{(j)}) $ to the agent variable node $\V{x}_{n} $ are calculated. For $s\neq s'$, the messages $\gamma^{(j)}_{ss'}(\V{x}_{n})$ propagating from the factor node $\underline{q}_{\mathrm{D}}(\V{x}_{n}, \underline{\V{y}}_{s,n}^{(j)},\underline{\V{y}}_{s',n}^{(j)}, \underline{\V{\beta}}_{ss',n}^{(j)}, \underline{a}_{ss',n}^{(j)};\V{z}_{n}^{(j)}) $ to the agent state variable node $\V{x}_{n} $ are calculated.

\subsubsection{Measurement Update for Legacy \acp{pmva}}
\label{sec:MeaUpdateLPVS}
For $s=s'$, the messages $\rho_{ss}(\underline{\V{y}}_{s,n}^{(j)}) = \rho_{ss}(\underline{\V{p}}_{s,\mathrm{\sfv}}^{(j)}, \underline{r}_{s,n}^{(j)}) $ propagating from the factor node $\underline{q}_{\mathrm{S}}(\V{x}_{n}, \underline{\V{y}}_{s,n}^{(j)}, \underline{\V{\beta}}_{ss,n}^{(j)}, \underline{a}_{ss,n}^{(j)};\V{z}_{n}^{(j)}) $ to the \ac{pmva} variable node $ \underline{\V{y}}_{s,n}^{(j)} $ are calculated. For $s\neq s'$, the messages $\rho_{ss'}(\underline{\V{y}}_{s,n}^{(j)}) = \rho_{ss'}(\underline{\V{p}}_{s,\mathrm{\sfv}}^{(j)}, \underline{r}_{s,n}^{(j)})$ sent to the \ac{pmva} variable node $\underline{\V{y}}_{s,n}^{(j)} $ are also calculated. With $\rho_{ss}(\underline{\V{y}}_{s,n}^{(j)})$ and $\rho_{ss'}(\underline{\V{y}}_{s,n}^{(j)})$, the messages $\gamma(\underline{\V{y}}_{s,n}^{(j)}) \triangleq \gamma(\underline{\V{p}}_{s,\mathrm{\sfv}}^{(j)}, \underline{r}_{s,n}^{(j)})$ sent to the next \ac{pa} are further calculated.

\subsubsection{Measurement Update for New \acp{pmva}}
The messages sent from $ \overline{q}_{\mathrm{N}}( \V{x}_{n}, \overline{\V{y}}_{m,n}^{(j)}, \overline{\V{\beta}}_{m,n}^{(j)}, \overline{a}_{m,n}^{(j)};\V{z}_{n}^{(j)}) $ to the variable node $\overline{\V{y}}_{m,n}^{(j)}$ are denoted as $\phi(\overline{\V{y}}_{m,n}^{(j)}) \triangleq \phi(\overline{\V{p}}_{m,\mathrm{\sfv}}^{(j)}, \overline{r}_{m,n}^{(j)}) $. 

\subsubsection{Measurement Update for Legacy \acp{pva}}
The messages $\kappa(\underline{\V{{\beta}}}^{(j)}_{00,n}) \triangleq \kappa(\underline{u}_{00,n}^{(j)}, \underline{r}_{00,n}^{(j)}) $ sent from the factor node $\underline{q}_{\mathrm{P}}(\V{x}_{n}, \underline{\V{\beta}}_{00,n}^{(j)}, \underline{a}_{00,n}^{(j)};\V{z}_{n}^{(j)})$ to the \ac{pva} node $\underline{\V{{\beta}}}^{(j)}_{00,n}$ are calculated. For $s=s'$, the messages sent from the factor node $\underline{q}_{\mathrm{S}}(\V{x}_{n}, \underline{\V{y}}_{s,n}^{(j)}, \underline{\V{\beta}}_{ss,n}^{(j)}, \underline{a}_{ss,n}^{(j)};\V{z}_{n}^{(j)}) $ to the \ac{pva} node $\underline{\V{{\beta}}}^{(j)}_{ss,n}$ are given as $\kappa(\underline{\V{{\beta}}}^{(j)}_{ss,n}) \triangleq \kappa(\underline{u}_{ss,n}^{(j)}, \underline{r}_{ss,n}^{(j)}) $. For $s\neq s'$, the messages $\kappa(\underline{\V{{\beta}}}^{(j)}_{ss',n}) \triangleq \kappa(\underline{u}_{ss',n}^{(j)}, \underline{r}_{ss',n}^{(j)}) $ sent from the factor node $\underline{q}_{\mathrm{D}}(\V{x}_{n}, \underline{\V{y}}_{s,n}^{(j)},\underline{\V{y}}_{s',n}^{(j)}, \underline{\V{{\beta}}}_{ss',n}^{(j)}, \underline{a}_{ss',n}^{(j)};\V{z}_{n}^{(j)}) $ to the \ac{pva} node $\underline{\V{{\beta}}}^{(j)}_{ss',n}$ are calculated. Based on the messages above, the messages $\gamma(\underline{\V{{\beta}}}^{(j)}_{ss',n}) \triangleq \gamma(\underline{u}_{ss',n}^{(j)}, \underline{r}_{ss',n}^{(j)}) $ sent to time $n+1$ are further obtained. 

\subsubsection{Measurement Update for New \acp{pva}}
The messages $\kappa(\overline{\V{{\beta}}}^{(j)}_{m,n}) \triangleq \kappa(\overline{u}_{mm,n}^{(j)}, \overline{r}_{mm,n}^{(j)}) $ sent from the factor node $ \overline{q}_{\mathrm{N}}( \V{x}_{n}, \overline{\V{y}}_{m,n}^{(j)}, \overline{\V{\beta}}_{m,n}^{(j)}, \overline{a}_{m,n}^{(j)};\V{z}_{n}^{(j)}) $ to the new \ac{va} variable node $\overline{\V{{\beta}}}^{(j)}_{m,n}$ are calculated.

\subsection{Belief Calculation}
After calculating the messages for all \acp{pa}, the belief $\tilde{f}(\V{x}_{n})$ of the agent state approximating $f(\V{x}_{n}|\V{z}_{1:n})$ is given by 
\begin{align}
	\tilde{f}(\V{x}_{n}) = C_{n}\alpha(\V{x}_{n}) \rrmv \prod_{ (s,s') \in \mathcal{D}_{n}^{(J)} } \rrmv \gamma^{(J)}_{ss'}(\V{x}_{n})  \prod_{j=1}^{J} \gamma^{(j)}_{00}(\V{x}_{n}) .
	\label{eq:BeliefAgent}
\end{align}
The beliefs of legacy \acp{pmva} $\tilde{f}(\underline{\V{y}}_{s,n}^{(J)}) \triangleq \tilde{f}(\underline{\V{p}}_{s,\mathrm{\sfv}}^{(J)}, \underline{r}_{s,n}^{(J)})$ with $s \in \mathcal{S}_{n}^{(J)}$ and of new \acp{pmva} $\tilde{f}(\overline{\V{y}}_{m,n}^{(J)}) \triangleq \tilde{f}(\overline{\V{p}}_{m,\mathrm{\sfv}}^{(J)}, \overline{r}_{m,n}^{(J)})$ with $m \in \mathcal{M}_{n}^{(J)}$ are given by
\begin{align}
	\tilde{f}(\underline{\V{y}}_{s,n}^{(J)}) & = \underline{C}_{s,n} \gamma(\underline{\V{y}}_{s,n}^{(J)}) \, ,
	\label{eq:BeliefLMVA} \vspace*{4mm}\\
	\tilde{f}(\overline{\V{y}}_{m,n}^{(J)}) & = \overline{C}_{m,n} \phi(\overline{\V{y}}_{m,n}^{(J)}) \, . 
	\label{eq:BeliefNewMVA}
\end{align}
At each PA $j$, the beliefs $\tilde{f}(\underline{\V{{\beta}}}^{(j)}_{ss',n}) \triangleq \tilde{f}(\underline{u}_{ss',n}^{(j)}, \underline{r}_{ss',n}^{(j)}) $ of legacy \acp{va} $ (s,s)'\in \tilde{\mathcal{D}}_{n}^{(j)} $ are given by 
\begin{align}
	\tilde{f}(\underline{u}_{ss',n}^{(j)}, \underline{r}_{ss',n}^{(j)}) = \underline{C}_{ss',n}^{(j)} \gamma(\underline{u}_{ss',n}^{(j)}, \underline{r}_{ss',n}^{(j)})
	\label{eq:BeliefLVA}
\end{align}
and the beliefs $\tilde{f}(\overline{\V{{\beta}}}^{(j)}_{m,n}) \triangleq \tilde{f}(\overline{u}_{mm,n}^{(j)}, \overline{r}_{mm,n}^{(j)}) $ of new \acp{pva} are given by
\begin{align}
	\tilde{f}(\overline{\V{{\beta}}}^{(j)}_{m,n}) = \overline{C}_{mm,n}^{(j)} \kappa(\overline{u}_{mm,n}^{(j)}, \overline{r}_{mm,n}^{(j)}).
	\label{eq:BeliefNewVA}
\end{align}
The normalization constants $C_{n}$, $\underline{C}_{s,n}$, $\overline{C}_{m,n}$, $\underline{C}_{ss',n}^{(j)}$, $\overline{C}_{mm,n}^{(j)}$ ensure that the beliefs are valid probability distributions, as an example, $\underline{C}_{s,n} = (\int \gamma(\underline{\V{p}}_{s,\mathrm{\sfv}}^{(J)}, 1)\mathrm{d} \underline{\V{p}}_{s,\mathrm{\sfv}}^{(J)} +  \gamma_{s,n}^{(J)} )^{-1}$.

\subsection{Implementation Aspect and Computational Complexity}
\label{subsec:ImpAsp}
As the integrations involved in the message and belief calculations cannot be obtained analytically, we employ a computationally efficient, sequential, particle-based, message-passing implementation, as outlined in \cite{Florian_TSP2017, LeiVenTeaMey:TSP2023, VenusTWC2024}. Specifically, we adopt a ``stacked state'' approach comprising the agent state, the \ac{pmva} states, and the \ac{pva} states, as detailed in \cite{MeyerTSPIN2016, VenusTWC2024}. This approach leads to complexity that scales linearly with the number of particles. The supplementary material of \cite{VenusTWC2024} provides a comprehensive description of the weight calculations when using ``stacked states''. \ac{spa} performed on loopy factor graph has been shown to converge and provides accurate approximations of marginal posterior \acp{pdf} \cite{Williams_loopyDA_2010, WilliamsLauTAE2014}. As the numbers of \acp{pmva} and their associated \acp{pva} grow with time, we remove \acp{pmva} with $ f( r_{s,n}^{(j)}=1 | \V{z}_{1:n}) $ below a threshold $p_{\mathrm{prun}}$, together with their associated \acp{pva} from the state space (``pruned'') at each \ac{pa} $j$ to limit computation complexity. Note that only the \acp{pva} with existence probabilities above $p_{\mathrm{prun}}$ are used for the measurement update of legacy \acp{pmva} in Section~\ref{sec:MeaUpdateLPVS}.

The conditional joint prior \ac{pdf} of each newly detected \ac{pmva} state and its associated \ac{pva} state is factorized as $f_{\mathrm{n}}(\overline{\V{p}}_{m,\mathrm{\sfv}}^{(j)}, \overline{u}_{mm,n}^{(j)}| \V{x}_{n}) = f_{\mathrm{n}}(\overline{\V{p}}_{m,\mathrm{\sfv}}^{(j)}| \V{x}_{n}) f_{\mathrm{n}}(\overline{u}_{mm,n}^{(j)})$. Instead of using the prior \acp{pdf} $f_{\mathrm{n}}(\overline{\V{p}}_{m,\mathrm{\sfv}}^{(j)}| \V{x}_{n}) $ and $f_{\mathrm{n}}(\overline{u}_{mm,n}^{(j)})$, we develop alternative proposal \acp{pdf} to account for the fact that a new surface could generate both single-bounce paths and double-bounce paths. As an example shown in Fig.~\ref{fig:GraphicalOverview_MVADemo}, the surface $3$ is solely supported by a double-bounce path until time step $126$ where a single-bounce path shows up. Note that conventional methods mostly use the single-bounce path assumption for initialization of new features. Early detection of new \acp{mva} supported by double-bounce paths enables the use of more measurements associated with these paths, enhancing both mapping and localization performance as shown in Section~\ref{subsec:synResults}. The states of new \acp{mva} and corresponding new \acp{va} are introduced considering following hypotheses, i.e., measurement $m$ can be generated by (i) the single-bounce path from a new \ac{pmva}; or (ii) a double-bounce path from the \ac{mva}-pair ``legacy \ac{pmva} $s \in \mathcal{S}_{n}^{(j)}$ to the new \ac{pmva}''; or (iii) \ac{mva}-pair ``new \ac{pmva} to the legacy \ac{pmva} $s \in \mathcal{S}_{n}^{(j)}$''. Specifically, for each measurement $\V{z}_{m,n}^{(j)}$, the initial states $\V{p}_{\mathrm{\sfv}}^{\mathrm{ini}}$ and $u^{\mathrm{ini}}$ are applied to $f_{\mathrm{S}}(\V{z}_{m,n}^{(j)}|\V{p}_{n}, \V{p}_{\mathrm{\sfv}}^{\mathrm{ini}}, u^{\mathrm{ini}})$ as in \eqref{eq:LHF_Spath} to calculate the \ac{lhf} for single-bounce hypothesis, and to $f_{\mathrm{D}}(\V{z}_{m,n}^{(j)}|\V{p}_{n}, \underline{\V{p}}_{s,\mathrm{\sfv}}^{(j)}, \V{p}_{\mathrm{\sfv}}^{\mathrm{ini}}, u^{\mathrm{ini}})$ and $f_{\mathrm{D}}(\V{z}_{m,n}^{(j)}|\V{p}_{n}, \V{p}_{\mathrm{\sfv}}^{\mathrm{ini}}, \underline{\V{p}}_{s,\mathrm{\sfv}}^{(j)}, u^{\mathrm{ini}})$ as in \eqref{eq:LHF_Dpath} for each legacy \ac{pmva} $s \in \mathcal{S}_{n}^{(j)}$ to calculate the \acp{lhf} for double-bounce hypotheses. The particles for the states $\V{p}_{\mathrm{\sfv}}^{\mathrm{ini}}$ and $u^{\mathrm{ini}}$ are obtained by sampling from the initial prior \acp{pdf} $f(\V{p}_{\mathrm{\sfv}}^{\mathrm{ini}}) $ and $f(u^{\mathrm{ini}}) $. The new \ac{pmva} state $\overline{\V{p}}_{m,\mathrm{\sfv}}^{(j)}$ and associated new \ac{pva} state $\overline{u}_{mm,n}^{(j)}$ needed for \eqref{eq:LHFNew} are further obtained by resampling the particles representing $\V{p}_{\mathrm{\sfv}}^{\mathrm{ini}}$ and $u^{\mathrm{ini}}$ based on the proposal \ac{pdf}, i.e., the aggregated \ac{lhf} $f_{\forall s}(\V{z}_{m,n}^{(j)}|\V{p}_{n}, \overline{\V{p}}_{m,\mathrm{\sfv}}^{(j)}, \underline{\V{p}}_{s,\mathrm{\sfv}}^{(j)}, \overline{u}_{mm,n}^{(j)})$ of all the above \acp{lhf}. Thus, the \ac{lhf} in \eqref{eq:LHFNew} is given as $f_{\mathrm{N}}(\V{z}_{m,n}^{(j)}|\V{p}_{n}, \overline{\V{p}}_{m,\mathrm{\sfv}}^{(j)}, \overline{u}_{mm,n}^{(j)}) \triangleq f_{\forall s}(\V{z}_{m,n}^{(j)}|\V{p}_{n},  \overline{\V{p}}_{m,\mathrm{\sfv}}^{(j)},  \underline{\V{p}}_{s,\mathrm{\sfv}}^{(j)}, \\ \overline{u}_{mm,n}^{(j)})$. Clearly, the double-bounce hypothesis requires a converged state $\underline{\V{p}}_{s,\mathrm{\sfv}}^{(j)}$ of the paired legacy \ac{pmva}. In general, mapping begins with a combination of single- and double-bounce paths for each surface, as environments containing only double- or higher-order bounce paths are rare. At the beginning, surfaces around the agent are rapidly identified primarily using single-bounce paths. Over time, their states gradually converge and are used to initialize new surfaces supported by double-bounce paths derived from them. 

Iterative message passing is performed for probabilistic \ac{da}. The computational complexity per \ac{da} iteration at the $j$th processing block is $\mathcal{O}(L_{n}^{(j)}M_{n})$ \cite{Florian_Proceeding2018, Erik_SLAM_TWC2019, LeiVenTeaMey:TSP2023}, where $L_{n}^{(j)} = \vert \tilde{\mathcal{D}}_{n}^{(j)} \vert = 1+ S_{n}^{(j)2}$ is the number of \emph{\ac{pmva}-oriented} association variables $ \underline{a}_{ss',n}^{(j)} $ and $M_{n}$ is the number of \emph{measurement-oriented} association variables $\overline{a}_{m,n}^{(j)}$. The number of iterations required to satisfy the convergence criterion is shown to be bounded \cite{WilliamsLauTAE2014}. With proper pruning at each processing block, the complexity is further reduced significantly. Moreover, by generating the proposal \ac{pdf} for new \acp{pmva} as mentioned above, we still follow the assumption ``one measurement--one new \ac{pmva}'' as in \cite{LeiVenTeaMey:TSP2023}, therefore the probabilistic \ac{da} is not further complicated due to considering multiple \ac{pmva}-pairs for each measurement.

%% file: InputFiles/ExperimentalResults.tex
The proposed method \ac{prop} is validated using both synthetic and real \ac{rf} measurements. First, using synthetic measurements from a \ac{dmimo} propagation scenario shown in Fig.~\ref{fig:GraphicalOverview_MVADemo}, the performance of \ac{prop} is validated and compared to two reference methods exploiting \acp{va}: (i) the \textit{\ac{refa}} from \cite{Erik_SLAM_TWC2019} extended with amplitude statistical model \cite{LeitingerICC2019, XuhongTWC2022}; (ii) the \textit{\ac{chslam}} from \cite{GentnerTWC2016} excluding the channel estimation stage. We extended \ac{refa} and \ac{chslam} to a \ac{mimo} setup considering angular statistical models, i.e., ranges, \acp{aod}, \acp{aoa} are considered for all paths in all methods. However, the reference methods apply the single-bounce assumption for all \acp{va} when calculating \acp{aod}, leading to mismatch for \acp{aod} of double-bounce paths in \acp{lhf}, therefore double-bounce measurements are mostly evaluated as \acp{fa} in \ac{refa} and \ac{chslam}. For consistency with \ac{prop} and \ac{refa} exploiting adaptive variances as shown in Section~\ref{subsec:MeaModel}, \ac{chslam} used Fisher information-based variances determined with amplitude measurements $ {z_\mathrm{u}}_{m,n}^{(j)}$. Note that both reference methods assume independent statistics across \acp{va}, hence they do not fuse information about map features across propagation paths like \ac{prop}.

\subsection{Simulation Setup}
\label{subsec:simsetup}
The following parameters are used for \ac{prop}, \ac{refa} and \ac{chslam} unless otherwise stated. The state-transition \ac{pdf} $f(\V{x}_{n}|\V{x}_{n-1}) $ of the agent state is defined by a linear near constant-velocity motion model\cite{BarShalom_AlgorithmHandbook}, i.e., $ \V{x}_{n} = \V{F}\V{x}_{n-1} + \V{\Gamma}\V{\nu}_{n} $, where the transition matrices $\V{F} \in \mathbb{R}^{4\times4} $ and $ \V{\Gamma} \in \mathbb{R}^{4\times2} $ are chosen as in \cite{BarShalom_AlgorithmHandbook} with the sampling period $ \Delta T $. The driving process $ \V{\nu}_{n} \in \mathbb{R}^{2\times1} $ is \ac{iid} across time $ n $, zero-mean and Gaussian with covariance matrix $ \sigma_{\mathrm{\nu}}^2 \mathbf{I}_{2} $, where $ \sigma_{\mathrm{\nu}} $ represents the average speed increment along the $x$ or $y$ axis during $ \Delta T $ and $ \mathbf{I}_{2} $ is a $2\times2$ identity matrix. The state-transition \ac{pdf} of the agent orientation is given as $ {\Delta\varphi}_{n} = {\Delta\varphi}_{n-1} + {\epsilon_{\mathrm{\varphi},n}} $ where the noise ${\epsilon_{\mathrm{\varphi},n}}$ is zero-mean and Gaussian with variance ${\sigma^2_{\mathrm{\varphi}}}$. The state-transition \ac{pdf} of the \ac{pva} normalized amplitude $\underline{u}_{ss',n}^{(j)}$ is chosen to be $ \underline{u}_{ss',n}^{(j)} = u_{ss',n-1}^{(j)} + {\epsilon_{ss',n}} $, where the noise $\epsilon_{ss',n}$ is \ac{iid} across $ (s\rmv,\rmv s') \in \tilde{\mathcal{D}}_{n}^{(j)} $, $j $ and $ n $, zero-mean, and Gaussian with variance ${\underline{\sigma}^2_{ss',n}}$. Geometrically, \acp{mva} represent static reflective surfaces and \acp{va} represent mirror images of static \acp{pa}, therefore their positions should remain unchanged over time. However, for the sake of numerical stability and to account for surface nonidealities such as slight curvature, a regularization noise is introduced in the state-transition \acp{pdf} of \acp{pmva} in \ac{prop}. The \acp{pdf} $f(\underline{\V{p}}_{s,\mathrm{\sfv}}|\V{p}_{s,\mathrm{\sfv}})$ and $f(\underline{\V{p}}_{s,\mathrm{\sfv}}^{(j)}|\V{p}_{s,\mathrm{\sfv}}^{(j-1)})$ in \eqref{eq:STpdf_MVA1} and \eqref{eq:STpdf_MVA3} are given as $\underline{\V{p}}_{s,\mathrm{\sfv}} = \V{p}_{s,\mathrm{\sfv}} + \V{\epsilon}_{s}$ and $\underline{\V{p}}_{s,\mathrm{\sfv}}^{(j)} = \V{p}_{s,\mathrm{\sfv}}^{(j-1)} + \V{\epsilon}_{s}$, where $\V{\epsilon}_{s}$ is \ac{iid} across $s$, zero-mean, and Gaussian with variance $ \sigma_{s}^2 \mathbf{I}_{2} $. Similarly, we also apply a Gaussian regularization noise in the state-transition \acp{pdf} of \ac{va} position states in \ac{refa} and \ac{chslam} with variance $ \sigma_{\mathrm{va}}^2 \mathbf{I}_{2} $. Furthermore, only the legacy \acp{pmva} that are consecutively detected over $3$ time steps are considered for the initialization of new \acp{pmva}.

The samples for the initial agent state are drawn from a $ 5 $-D uniform distribution centered at $ [\V{p}_{0}^{\mathrm{T}}\iist 0 \iist 0 \iist {\Delta\varphi}_{0}]^{\mathrm{T}} $ where $ \V{p}_{0} $ and ${\Delta\varphi}_{0}$ are the true agent start position and orientation, and the support of position, velocity and orientation components are $ [-0.5,0.5]\,$m, $ [0.01,0.01]\,$m/s and $ [-10^{\circ},10^{\circ}]$, respectively. The samples representing initial state of new \acp{pmva} $\V{p}_{\mathrm{\sfv}}^{\mathrm{ini}}$ are drawn from the \ac{pdf} $f(\V{p}_{\mathrm{\sfv}}^{\mathrm{ini}}) $ which is uniform on the \ac{roi}, i.e., $[-15\mathrm{m},15\mathrm{m}] \times[-15\mathrm{m},15\mathrm{m}] $ centered at the coordinate center. The samples representing the initial normalized amplitude state $u^{\mathrm{ini}}$ are drawn from a Gaussian distribution with mean ${z_\mathrm{u}}_{m,n}^{(j)}$ and variance calculated using the measurement $ {z_\mathrm{u}}_{m,n} $. We performed $100$ simulation runs with the following parameters: the mean number of \acp{fa} $\mu_{\mathrm{fa}} = 2$, the mean number of new \acp{pmva} $\mu_{\mathrm{n}} = 0.1$ in \ac{prop}, the mean number of new \acp{va} $\mu_{\mathrm{n}}^{\mathrm{va}} = 0.1$ in \ac{refa}, the detection threshold for normalized amplitude measurements $ u_{\mathrm{de}} = 6\,$dB, the probability of survival $p_{\mathrm{s}} = 0.99$, the pruning threshold $p_\mathrm{prun} = 0.1$, the existence detection thresholds for \acp{pva} and \acp{pmva} $ p_\mathrm{\pr} =0.5 $ and $ p_\mathrm{\sfv} =0.5 $, $ 60000 $ particles for each random variable state, the agent acceleration noise variance $ \sigma_{\mathrm{\nu}}^2 = 9\cdot10^{-4}\,$$ \mathrm{m}/\mathrm{s}^2 $, the orientation noise standard deviation ${\sigma_{\mathrm{\varphi}}} = 7^{\circ}$, the regularization noise standard deviations for \ac{pmva} and \ac{va} position states $\sigma_{s} = 0.01\,$m and $\sigma_{\mathrm{va}} = 0.01\,$m, the normalized amplitude noise standard deviation $ {\underline{\sigma}_{ss',n}} = 0.02\, \hat{u}_{ss',n-1}^{(j)} $.\footnote{The heuristic approach to scale the standard deviation ${\underline{\sigma}_{ss',n}} $ by the \ac{mmse} estimates was chosen considering that the range of normalized amplitude values tends to be very large.} For \ac{chslam}, we used $2000$ particles for the agent and $1000$ for the \acp{va}.

The \ac{fa} measurements originating from the pre-processing stage are modeled by a Poisson point process with mean number $ \mu_{\mathrm{fa}} $ and \ac{pdf} factorized as $ f_{\mathrm{fa}}(\V{z}_{m,n}^{(j)}) =  f_{\mathrm{fa}}({z_\mathrm{d}}_{m,n}^{(j)}) f_{\mathrm{fa}}({z_\mathrm{\phi}}_{m,n}^{(j)}) f_{\mathrm{fa}}({z_\mathrm{\varphi}}_{m,n}^{(j)}) f_{\mathrm{fa}}({z_\mathrm{u}}_{m,n}^{(j)} )$, where the individual \acp{pdf} corresponding to distances, \acp{aod} and \acp{aoa} are uniform on $ [0,d_{\mathrm{max}}] $, $ [-\pi,\pi) $ and $ [-\pi,\pi) $, respectively. For the normalized amplitudes, $ f_{\mathrm{fa}}({z_\mathrm{u}}_{m,n}^{(j)} ) $ is given by a truncated Rayleigh \ac{pdf} with $ 1/2 $ scale parameter \cite{XuhongTWC2022, AlexTWC2024}. 

\subsection{Performance Metrics}
\label{subsec:Metric}

The performance of different methods is measured in terms of their \acp{rmse} of the agent position and orientation, and the \ac{ospa} error \cite{OSPA_TSP2008} of \acp{va} and \acp{mva}\footnote{Note that the \ac{ospa} error is normalized by the larger cardinality of the comparing sets. Correspondingly, we normalize the error bound with the true number of \acp{mva}.} with cutoff distance $5\,$m and order $1$. The \ac{ospa} error is a distance metric that quantifies the difference between a set of estimates and known truths, and it accounts for both the estimation error and cardinality error. In addition, we compute the \acp{pcrlb} \cite{Tichavsky_PosteriorCRLB_1998, FasDeuKesWilColWitLeiSecWym:STSP2025, DeutschmannAsilomar25} as a performance benchmark, which explicitly characterizes the localization and mapping error bounds while accounting for data fusion across multiple \acp{pa} and propagation paths, as well as system-related factors like array geometry, \ac{snr}, and bandwidth.
The mean \acp{rmse}, \ac{mospa} errors are obtained at each time by averaging over all converged simulation runs and compared against the \ac{peb}, \ac{oeb}, and \acp{meb} derived in~\cite{DeutschmannAsilomar25}. 
Furthermore, to illustrate the contribution of different types of propagation paths to the achievable performance, error bounds are reported for three different setups: (i) DB, e.g., \ac{peb}(DB), the \ac{peb} considering \acp{los}, single and double-bounce paths. (ii) (SB), the bound considering \acp{los}, single-bounce paths and (iii) (LoS), the bound considering \acp{los} only. A simulation run with $\forall n: \norm{\hat{\V{x}}_{n} - \V{x}_{n}} <1\,$m and $\forall s: \norm{\hat{\V{p}}_{s,\mathrm{sfv}} - \V{p}_{s,\mathrm{sfv}}} <1\,$m is considered as converged.

\begin{figure}[!t]
	\centering
	\hspace*{7.7mm}\subfloat[\label{subfig:synResultPA1fig1}]
	{\hspace*{-9.7mm}\includegraphics[scale=0.94]{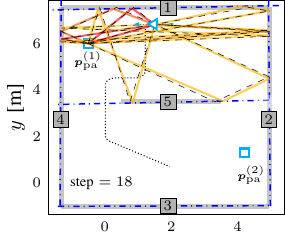}}
	\hspace*{5.5mm}\subfloat[\label{subfig:synResultPA2fig1}]
	{\hspace*{-5.5mm}\includegraphics[scale=0.94]{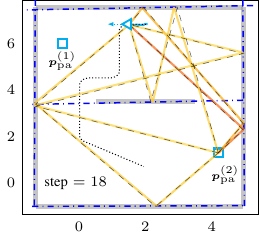}}\\[-2.8mm]
	\hspace*{7.7mm}\subfloat[\label{subfig:synResultPA1fig3}]
	{\hspace*{-9.7mm}\includegraphics[scale=0.94]{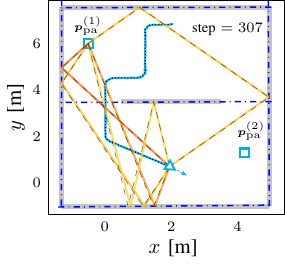}}
	\hspace*{5.5mm}\subfloat[\label{subfig:synResultPA3fig3}]
	{\hspace*{-5.5mm}\includegraphics[scale=0.94]{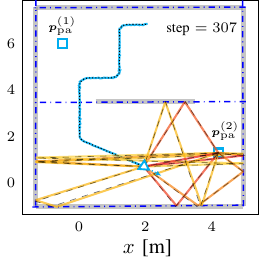}}\\[-1.8mm]
	{\makebox[0.5\textwidth]
		{\hspace*{3.65mm}\includegraphics[scale=0.8978]{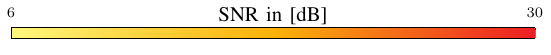}} 
	}\quad \\[-2.55mm]
	{\makebox[0.5\textwidth]
		{\hspace*{3mm}\includegraphics[scale=0.98]{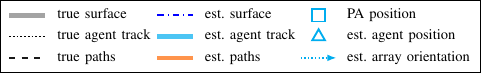}} 
	}\quad \\[0mm]
	\caption{Results of a simulation run of \ac{prop} for Experiment~$1$, showing the true and estimated reflective surfaces, propagation paths, and agent positions at time steps $n = 18$ in (a), (b) and $n = 307$ in (c), (d), respectively. In (a), the surfaces are labeled in the same order as in Fig.~\ref{fig:GraphicalOverview_MVADemo}. The line representations of the estimated surfaces are computed from the \ac{mmse} estimates of the detected \acp{mva}. Estimated propagation paths are obtained by connecting the \ac{mmse} estimate of the agent position, the interaction points on the estimated surfaces, and the \acp{pa}, and are compared with the true visible paths at each time step. The color of each estimated path represents its estimated \ac{snr}, i.e., the square of the normalized amplitude \ac{mmse} estimate.}	 
	\label{fig:synResultTwoPAs}
	\vspace{-1mm}
\end{figure}

\subsection{Synthetic Measurements}
Measurements $ \V{z}_{m,n}^{(j)} $ are synthetically generated according to the scenario shown in Fig.~\ref{fig:GraphicalOverview_MVADemo}, where a \ac{dmimo} system operating at $f_{\mathrm{c}} = 6$\,GHz with bandwidth of $1$\,GHz is used. The same $ 5 \times 5 $ uniform rectangular array with inter-element spacing of $\lambda/4$ is used at both \acp{pa} and the mobile agent. Over time steps, propagation paths experiencing up to double bounces with time-varying distances, \acp{aoa}, \acp{aod}, and normalized amplitudes were synthesized. We assume that the \ac{pa} orientations are $0^{\circ}$ and the true agent orientation is consistent with the direction of movement. The amplitude of each path is assumed to follow free-space path loss and additionally attenuated by $3$\,dB after each bounce on a surface. The output \ac{snr} at $1$\,m from the \ac{pa} is assumed to be $\mathrm{SNR}_{\mathrm{1m}}=30$\,dB, according to which the measurement noise standard deviation is further calculated based on the Fisher information \cite{Thomas_Asilomar2018, LeitingerICC2019, XuhongTWC2022}. In each simulation run, noisy measurements are generated according to \eqref{eq:LHF_dist}--\eqref{eq:LHF_normAmp}, i.e., adding noises (determined based on the Fisher information) to the true path parameters, and stacked into the vector $\V{z}_{n}^{(j)}$\!. In addition, \ac{fa} measurements are generated with mean number $ \mu_{\mathrm{fa}}=2 $ and added to $\V{z}_{n}^{(j)}$\!. In the following, we show results for two synthetic measurement setups: (i) \textit{Experiment~1}: both \ac{pa}\,$1$ and \ac{pa}\,$2$ are used, the agent moves from point A to C in Fig.~\ref{fig:GraphicalOverview_MVADemo}, leading to a
total of $307$ time steps. \ac{los} propagation applies to at least one \ac{pa}; (ii) \textit{Experiment~2}: only \ac{pa}\,$1$ is used, the agent moves from point B to D, leading to a total of $312$ time steps. \Ac{olos} condition applies after time step $220$.

\begin{figure}[!t]
	\centering
	\hspace*{6mm}\subfloat
	{\hspace*{0.7mm}\includegraphics[scale=0.912]{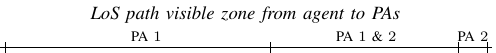}}\\[-3mm]
	\addtocounter{subfigure}{-1}
	\hspace*{4.7mm}\subfloat[\label{subfig:visibilityAnalysis}]
	{\hspace*{-6mm}\includegraphics[scale=0.915]{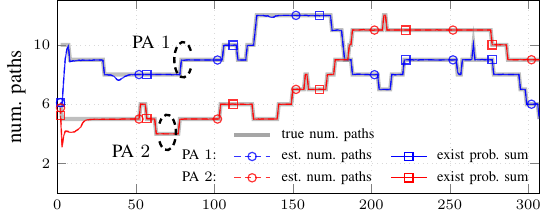}}\\[-2.5mm]
	\hspace*{5.5mm}\subfloat[\label{subfig:mospa_distance_VA}]
	{\hspace*{-6mm}\includegraphics[scale=0.915]{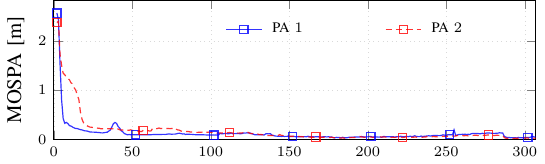}}\\[-2.5mm]
	\hspace*{6mm}\subfloat[\label{subfig:mospa_distance_MVA}]
	{\hspace*{-8mm}\includegraphics[scale=0.915]{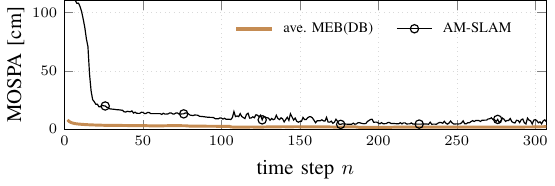}}\\[-1mm]
	\caption{Performance results of ``soft'' ray tracing using \ac{prop} for Experiment~$1$. (a) For each \ac{pa}, the true number of visible paths, the number of detected paths, and the sum of the estimated existence probabilities of the detected paths are shown. (b) \ac{mospa} errors of the estimated \acp{va}, obtained by geometric transformation of the \ac{mva} estimates at each \ac{pa}. (c) \ac{mospa} errors of the estimated \acp{mva}, including a comparison with the \acp{meb}(DB), averaged over all surfaces. All values are averaged over all simulation runs at each time step.}	 
	\label{fig:pathTracingResult}
\end{figure}

\begin{figure*}[!t]
	\centering
	\hspace*{5mm}\subfloat[\label{subfig:MSEsBoundS1sum}]
	{\hspace*{-6mm}\includegraphics[scale=1]{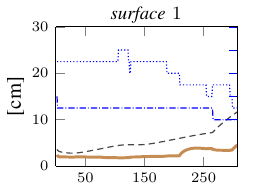}}
	\hspace*{-2mm}\subfloat[\label{subfig:MSEsBoundS2sum}]
	{\hspace*{-0mm}\includegraphics[]{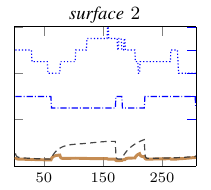}}
	\hspace*{-2.3mm}\subfloat[\label{subfig:MSEsBoundS3sum}]
	{\hspace*{0.3mm}\includegraphics[]{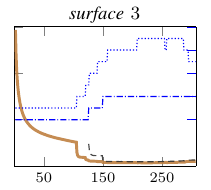}}
	\hspace*{-2.3mm}\subfloat[\label{subfig:MSEsBoundS4sum}]
	{\hspace*{0.3mm}\includegraphics[]{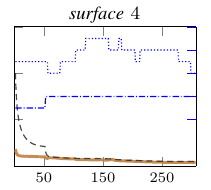}}
	\hspace*{-7.2mm}\subfloat[\label{subfig:MSEsBoundS5sum}]
	{\hspace*{5.2mm}\includegraphics[]{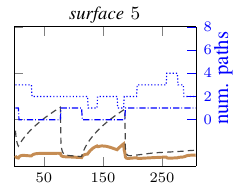}}\\[-4mm]
	\hspace*{0.7mm}\subfloat[\label{subfig:MSEsBoundS1}]
	{\hspace*{-8.5mm}\includegraphics[scale=1]{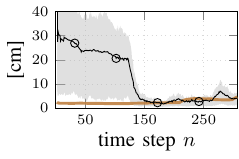}}
	\hspace*{2mm}\subfloat[\label{subfig:MSEsBoundS2}]
	{\hspace*{-1.34mm}\includegraphics[]{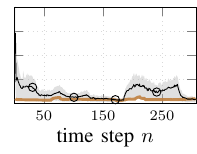}}
	\hspace*{2.1mm}\subfloat[\label{subfig:MSEsBoundS3}]
	{\hspace*{-1.34mm}\includegraphics[]{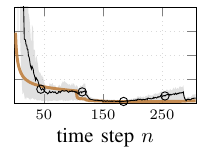}}
	\hspace*{1.5mm}\subfloat[\label{subfig:MSEsBoundS4}]
	{\hspace*{-1.34mm}\includegraphics[]{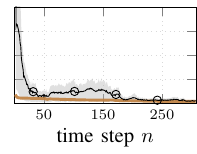}}
	\hspace*{1.5mm}\subfloat[\label{subfig:MSEsBoundS5}]
	{\hspace*{-1.34mm}\includegraphics[]{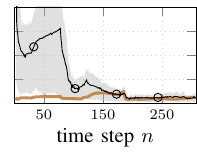}}\\[-2mm]
	\hspace*{2mm}\subfloat
	{\hspace*{-2.5mm}\includegraphics[]{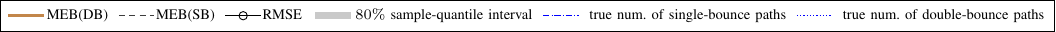}}\\[-1mm]
	\caption{Performance results of \ac{prop} for Experiment~$1$. For each surface labeled as $s \in \{1,\dots,5\}$ in Figs.~\ref{fig:GraphicalOverview_MVADemo} and \ref{subfig:synResultPA1fig1}, (a)–(e) show the \acp{meb} and the corresponding true numbers of single- and double-bounce paths over time, while (f)–(j) show the \acp{rmse} of the \ac{mva} position estimates. The shaded bands around the mean \acp{rmse} indicate the variability of the error samples, corresponding to central $80\%$ quantile intervals \cite{DeutschmannAsilomar25}.}	
	\label{fig:MEBperSurface}
\end{figure*}

\subsection{Results for Synthetic Measurements}
\label{subsec:synResults}
Fig.~\ref{fig:synResultTwoPAs} depicts for one exemplary simulation run of \ac{prop} the \ac{mmse} estimates of the agent positions and orientations, the detected reflective surfaces (obtained from the \ac{mmse} estimates of \ac{mva} positions), as well as the detected \acp{pva}. It is shown that \ac{prop} performs accurate ray tracing, the paths at each \ac{pa} are accurately detected and estimated in terms of their time-varying geometrical parameters and \acp{snr}. Leveraging all \acp{pa} and visible \acp{mpc}, \ac{prop} accurately localized the agent and rapidly identified all five surfaces including the lower one (surface $3$) supported by a single double-bounce path until steps $n>126$, proving the mapping capability to bypass \ac{olos} conditions. Fig.~\ref{fig:pathTracingResult} illustrates more on the ``soft'' ray tracing. It can be seen that \ac{prop} accurately adapts to the time-varying path visibilities and states at all \acp{pa}. Specifically, as shown in Fig.~\ref{subfig:visibilityAnalysis} the numbers of detected paths rapidly converge to the true path numbers, and in Fig.~\ref{subfig:mospa_distance_VA} the \acp{mospa} errors of the \ac{va} positions rapidly drop below $20\,$cm at the beginning. The close alignment between the sum of the estimated path existence probabilities and the true number of paths further indicates the strong confidence in the estimates, with individual existence probabilities approaching one.

In the following, we present the statistical performance evaluation\footnote{We excluded outliers (unless stated otherwise) with values exceeding $1.5$ interquartile ranges above the upper quartile or below the lower quartile \cite{DeutschmannAsilomar25}.}. For both experiments, none of the $100$ simulation runs diverged for \ac{prop} and \ac{refa}. For \ac{chslam}, $5$ and $26$ runs diverged in Experiment~1 and Experiment~2, respectively\footnote{In Experiment $ 2 $, for \ac{refa} and \ac{chslam}, we only evaluate the estimates up to step $ 285 $ for the convergence check, since beyond this step only double-bounce paths exist and all runs for the two methods diverged.}. All simulations are conducted on an AMD Ryzen $9$ $5950$X CPU ($16$ cores, $32$ threads; $2.2-3.4$\,GHz). The average runtimes per time step (MATLAB implementations) are $2.54$\,s for \ac{prop}, $0.6$\,s for \ac{refa}, and $9.43$\,s for \ac{chslam}.

\begin{figure*}[!t]
	\centering
	\hspace*{0mm}\subfloat
	{\hspace*{7mm}\includegraphics[scale=1]{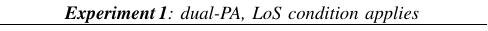}}
	\hspace*{0mm}\subfloat
	{\hspace*{8mm}\includegraphics[]{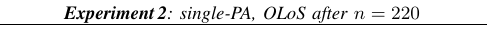}}\\[-4mm]
	\addtocounter{subfigure}{-2}
	\hspace*{6mm}\subfloat[\label{subfig:OEB_agentOrientation_s1}]
	{\hspace*{-8mm}\includegraphics[scale=1]{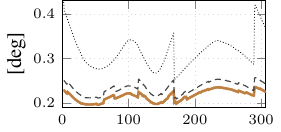}}
	\hspace*{6mm}\subfloat[\label{subfig:PEB_agentPosition_s1}]
	{\hspace*{-8mm}\includegraphics[]{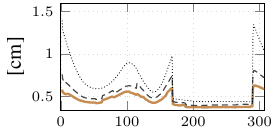}}
	\hspace*{7mm}\subfloat[\label{subfig:OEB_agentOrientation_s2}]
	{\hspace*{-8mm}\includegraphics[scale=1]{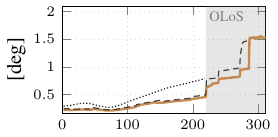}}
	\hspace*{5mm}\subfloat[\label{subfig:PEB_agentPosition_s2}]
	{\hspace*{-7mm}\includegraphics[]{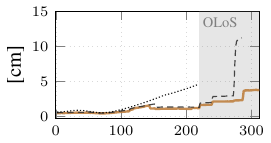}}\\[-4mm]
	\hspace*{6mm}\subfloat[\label{subfig:mospa_agentOrientation_s1}]
	{\hspace*{-7mm}\includegraphics[scale=1]{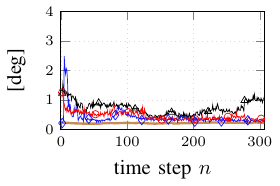}}
	\hspace*{5mm}\subfloat[\label{subfig:mospa_agentPosition_s1}]
	{\hspace*{-7mm}\includegraphics[]{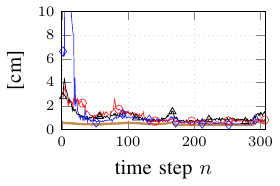}}
	\hspace*{6.5mm}\subfloat[\label{subfig:mospa_agentOrientation_s2}]
	{\hspace*{-5mm}\includegraphics[scale=1]{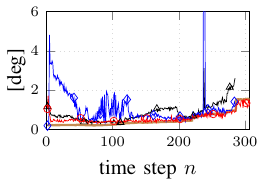}}
	\hspace*{4mm}\subfloat[\label{subfig:mospa_agentPosition_s2}]
	{\hspace*{-6.6mm}\includegraphics[]{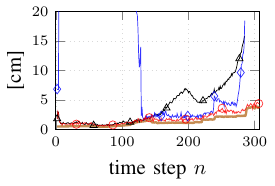}}\\[-4mm]
	\hspace*{7mm}\subfloat[\label{subfig:PEB_CDF_scenario1}]
	{\hspace*{-8mm}\includegraphics[scale=1]{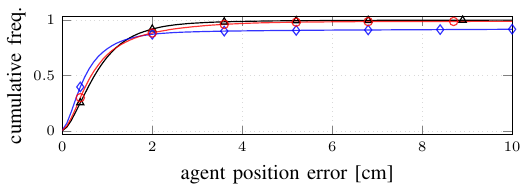}}
	\hspace*{13mm}\subfloat[\label{subfig:PEB_CDF_scenario2}]
	{\hspace*{-13mm}\includegraphics[]{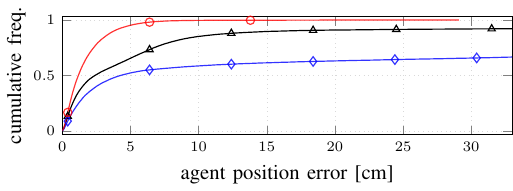}}\\[-2.5mm]
	\hspace*{-2mm}\subfloat
	{\hspace*{2mm}\includegraphics[]{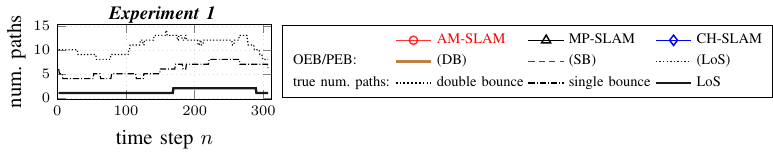}}
	\hspace*{-2mm}\subfloat
	{\hspace*{2.5mm}\includegraphics[]{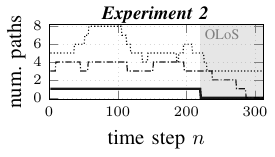}}\\[-1mm]
	\caption{Performance results of \ac{prop}, \ac{refa}, and \ac{chslam} for Experiments~$1$ and $2$ with synthetic measurements. (a)–(d) \acp{oeb} and \acp{peb} for different path sets, i.e., LoS, single-bounce (SB), and double-bounce (DB). (e)–(h) \acp{rmse} of the agent orientation and position, including a comparison with the corresponding \acp{oeb}(DB) and \acp{peb}(DB). (i)–(j) Cumulative frequencies of the \acp{rmse} (including outliers) of the agent position. In addition, the true numbers of LoS, single-bounce, and double-bounce paths over time are shown for both experiments.}	 
	\label{fig:result1}
\end{figure*}

Figs.~\ref{subfig:MSEsBoundS1sum} to \ref{subfig:MSEsBoundS5sum} show the \acp{meb}(SB) and \acp{meb}(DB) introduced in \cite{DeutschmannAsilomar25}, where the gap between the bounds for each surface highlights the importance of incorporating double- or higher order-bounce paths for mapping, particularly under \ac{olos} conditions and when single-bounce paths are unavailable. It is also shown that the \acp{meb} effectively capture the impacts of individual paths. For instance, for surface $2$ the rise in \ac{meb}(DB) around time step $n=220$ indicates a brief disappearance of a double-bounce path. Figs.~\ref{subfig:MSEsBoundS1} to \ref{subfig:MSEsBoundS5} further present the \acp{rmse} of \acp{mva} positions and compared against the \acp{meb}(DB), where the associations between the \acp{mva}' estimates and ground truth were decided using the Hungarian method \cite{HungarianArticle1955}. In Figs.~\ref{subfig:MSEsBoundS1} and \ref{subfig:MSEsBoundS5}, slightly higher \acp{rmse} are observed for surfaces $1$ and $5$ at the beginning, which arise from the agent's initial orientation ambiguity leading to the first established surfaces to become stuck in tilted states. This is resolved once the agent undergoes sufficient directional change, specifically after the agent's first turn at around step $30$, the \acp{rmse} start to drop to the \ac{meb}(DB) level. Overall, \ac{prop} converges asymptotically to the \acp{meb}(DB) for individual surfaces. As also depicted in Fig.~\ref{subfig:mospa_distance_MVA}, the mean \acp{mospa} of the detected \acp{pmva} rapidly attain the \acp{meb} averaged over all surfaces. 

Fig.~\ref{fig:result1} shows for \ac{prop}, \ac{refa} and \ac{chslam} the \acp{rmse} of agent orientation and position, and the comparison against the \ac{oeb}(DB) and \ac{peb}(DB), respectively. As shown in Figs.~\ref{subfig:mospa_agentOrientation_s1} and \ref{subfig:mospa_agentPosition_s1} for Experiment~1, as the map is rapidly established at the beginning, \ac{prop} efficiently leverages the information from both \acp{pa} and all single- and double-bounce paths, attains the \ac{oeb}(DB) and \ac{peb}(DB) with \acp{rmse} quickly dropping below $1^{\circ}$ and $1$\,cm. Note that \ac{refa} and \ac{chslam} also converge asymptotically to the \ac{peb}(DB) even though they exploit only \ac{los} and single-bounce paths, which is due to the contribution of higher-order bounce paths to localization is marginal under \ac{mimo} conditions with strong \acp{los}, as evidenced by the theoretical error bounds in Fig.~\ref{subfig:PEB_agentPosition_s1}. In Experiment~2 with the single-\ac{pa} setup, Figs.~\ref{subfig:mospa_agentOrientation_s2} and \ref{subfig:mospa_agentPosition_s2} show that \ac{prop} remains tightly aligned with the bounds from the beginning and significantly outperforms \ac{refa} and \ac{chslam} under \ac{olos} condition after time step $n=220$. Even beyond $n=286$ where only double-bounce paths remain and both \ac{refa} and \ac{chslam} completely fail, \ac{prop} still localizes accurately and stays close to the bounds. Figs.~\ref{subfig:PEB_CDF_scenario1} and ~\ref{subfig:PEB_CDF_scenario2} show the cumulative frequencies of the agent errors, including the diverged runs and outliers. The results show that \ac{prop} achieves statistically significant improvements in both accuracy and robustness (fewer instances
of large errors), especially in challenging \ac{olos} scenarios, highlighting the performance gain from fusion with double-bounce paths and incorporating amplitude information in the statistical models.

\begin{figure}[t!]
	\centering
	\includegraphics[width=8.35cm,height=4.55cm]{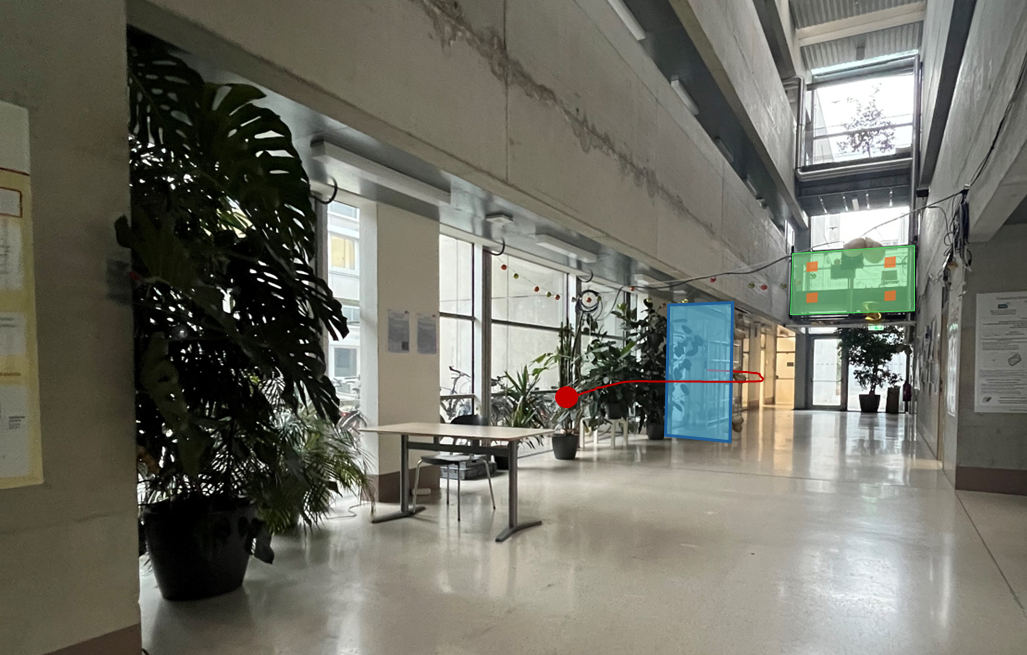}
	\caption{Overview of the hallway measurement environment at TU Graz, Austria. Ground-truth \acp{pa} positions and the agent trajectory are shown by brown rectangles and a solid red line, respectively.}
	\label{fig:hallWayPicture}
	\vspace{0mm}
\end{figure}

\begin{figure*}[!t]
	\centering
	\hspace*{0mm}\subfloat[\label{subfig:SchematicDepictionHallWay}]
	{\hspace*{-14mm}\includegraphics[scale=1]{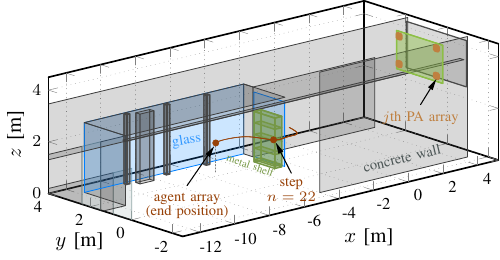}}
	\hspace*{14.5mm}\subfloat[\label{subfig:SchematicDepictionHallWay2D}]
	{\hspace*{-9.5mm}\includegraphics[]{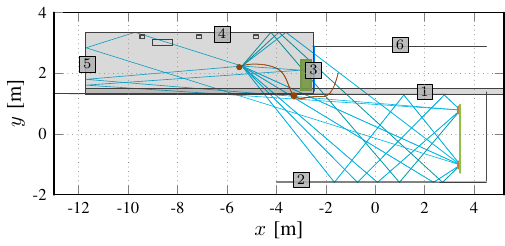}}\\[-3mm]
	\hspace*{7mm}\subfloat[\label{subfig:HallWayEstSurface}]
	{\hspace*{-9.5mm}\includegraphics[scale=1]{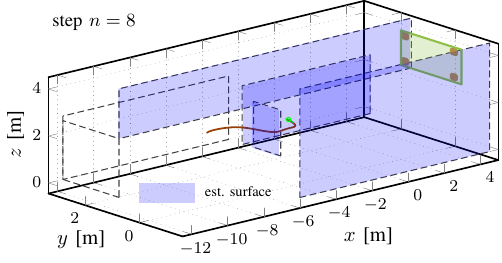}}
	\hspace*{1mm}\subfloat[\label{subfig:HallWay2DEstRays}]
	{\hspace*{1.5mm}\includegraphics[]{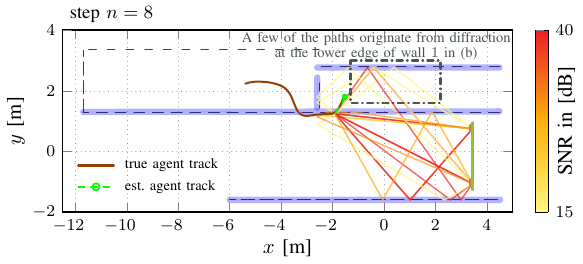}}\\[-3mm]
	\hspace*{7mm}\subfloat[\label{subfig:HallWayEstSurfaceEndPoint}]
	{\hspace*{-9.5mm}\includegraphics[scale=1]{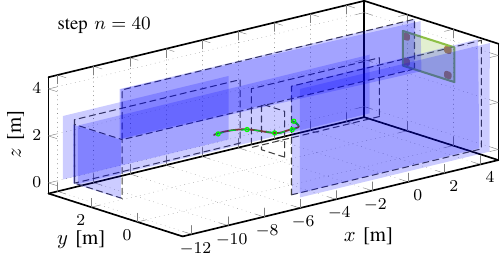}}
	\hspace*{1mm}\subfloat[\label{subfig:HallWay2DEstRaysEndPoint}]
	{\hspace*{1.5mm}\includegraphics[]{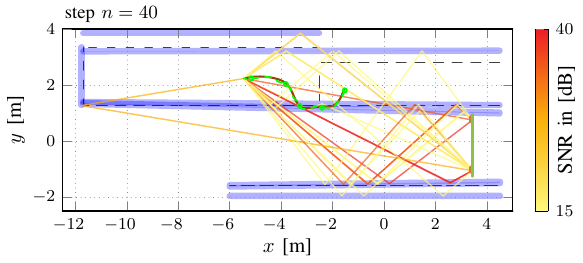}}\\[-1mm]
	\caption{Real measurement scenario and performance of \ac{prop}. (a) \ac{3d} schematic illustration of the hallway scenario (cf.~Fig.~\ref{fig:hallWayPicture}) and the measurement setup. (b) Top view of the environment and the propagation paths up to double bounces generated by ray tracing based on the floor plan at step $n = 40$. Six labeled surfaces indicate the main reflective surfaces in the environment. (c) and (e) Localization and mapping results (detected surfaces) in \ac{3d} at steps $n = 8$ and $n = 40$, respectively. (d) and (f) Top views of the corresponding ``soft'' ray-tracing results. Since \ac{prop} does not estimate surface dimensions, the estimated surfaces are shown using the true surface dimensions solely for visualization purposes.}	 
	\label{fig:hallWayDemoandResult}
	\vspace{-1mm}
\end{figure*}

The low mapping and agent \acp{rmse} of \ac{prop} demonstrate the great potential of \ac{prop} for accurate and robust sensing. Moreover, the mapping and ray tracing results suggest the broader applicability of the proposed method beyond mapping and localization. \ac{prop} infers the number of visible paths which upper-bounds the rank of \ac{mimo} channel matrix and hence the number of independent spatial data streams that can be transmitted over the \ac{rf} propagation channel. Moreover, it can effectively acquire and predict environment-specific channel state information, and assist applications like \ac{olos} beamforming and beamalignment via \acp{mpc}.

\subsection{Results for Real Measurements}
\label{subsec:RealResults}

We further validate the proposed method using real distributed \ac{mimo} synthetic aperture measurements acquired in a hallway scenario at TU Graz, Austria. The environment features multiple concrete and metal-coated glass walls as the primary reflective surfaces, along with rich scattering objects such as metal structures. More details about the measurement campaign can be found in \cite{BenjaminChannelPrediction_2025}. The channel coefficients between a mobile agent and $ J\,=\,4 $ distributed \acp{pa} were measured with a Rohde~\&~Schwarz ZVA$24$ vector network analyzer over a bandwidth of $500$\,MHz centered around $6.25$\,GHz. As shown in Fig.~\ref{fig:hallWayPicture} and the corresponding schematic depiction in Fig.~\ref{fig:hallWayDemoandResult}, the \acp{pa} were mounted on a bridge between two concrete walls on the second floor, and each \ac{pa} was equipped with a $4\times4$ uniform rectangular array. The agent was equipped with a $3\times3$ uniform rectangular array moved along a trajectory ($ 40 $ discrete time steps) around a glass wall and metal shelf, from \ac{los} to \ac{olos} conditions (after step $n = 22$). To simulate a strong \ac{los} obstruction in the latter half of the trajectory, the shelf is purposely filled with pyramidal \ac{rf} absorbers. In \ac{los} conditions, the measurements exhibit an input \ac{snr} of approximate $4$\,dB. Note that the arrays on both agent and \ac{pa} sides were synthetically formed using mechanical positioners. Accordingly, we extended the proposed algorithm to a \ac{3d} uplink formulation with both horizontal and vertical propagation from the agent to the \acp{pa}.

At each time step and at each \ac{pa}, we apply a variational sparse Bayesian parametric channel estimation algorithm~\cite{grebien2024SBL,MoePerWitLei:TSP2024} with detection threshold $u_{\mathrm{de}} = 15\,$dB to obtain ${M}_{n}^{(j)}$ measurement vectors $\V{z}_{m,n}^{(j)}$, consisting of the distance measurement ${\rv{z}_\mathrm{d}}_{m,n}^{(j)}$, \ac{aod} measurements in azimuth $ {\rv{z}_\mathrm{\phi}}_{m,n}^{(j)} $ and elevation $ {\rv{z}_\mathrm{\theta}}_{m,n}^{(j)} $, \ac{aoa} measurements in azimuth ${\rv{z}_\mathrm{\varphi}}_{m,n}^{(j)}$ and elevation ${\rv{z}_\mathrm{\vartheta}}_{m,n}^{(j)}$, and the normalized amplitude measurement ${\rv{z}_\mathrm{u}}_{m,n}^{(j)}\rmv\in\rmv [u_{\mathrm{de}}, \infty) $. 
For the real data, we used the parameter settings as in Section~\ref{subsec:simsetup}, except for the following: the orientation noise standard deviation ${\sigma_{\mathrm{\varphi}}} = 9^{\circ}$, the regularization noise standard deviations for \ac{pmva} $\sigma_{s} = 2\,$mm and the normalized amplitude noise standard deviation $ {\underline{\sigma}_{ss',n}} = 0.01\, \hat{u}_{ss',n-1}^{(j)} $.

\begin{figure}[!t]
	\captionsetup[subfigure]{labelformat=empty, labelsep=none} 
	\centering
	\vspace*{-4mm}
	\hspace*{6mm}\subfloat[]
	{\hspace*{-6mm}\includegraphics[scale=1]{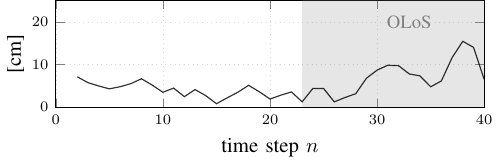}}\\[-1.5mm]
	\caption{Performance results of the proposed \ac{prop} for real measurements. Agent position estimation error over time.}
	\label{fig:realDataAgent}
\end{figure}

Fig.~\ref{fig:hallWayDemoandResult} depicts for one estimation run of \ac{prop} the estimated agent tracks, the detected reflective surfaces (obtained from the estimates of \ac{mva} positions), as well as the detected \acp{pva}. It is shown that \ac{prop} quickly establishes the map at the beginning of the trajectory, and identifies all six main reflective surfaces as labeled in Fig.~\ref{subfig:SchematicDepictionHallWay2D}, including surface~$3$ which is only visible at the beginning of the trajectory. Surface~$1$ is successfully estimated although it is visible only via double-bounce paths which particularly aid positioning in \ac{olos} conditions. ``Ghost'' surfaces detected alongside surfaces $4$ and $2$ arise from complex wall structures (as shown in Fig.~\ref{fig:hallWayPicture}) that introduce cluttered components in the channel measurements, resulting in short-lived detections of features that are not geometrically meaningful. Figs.~\ref{subfig:HallWay2DEstRays} and \ref{subfig:HallWay2DEstRaysEndPoint} further show the ``soft'' ray tracing results. The detected paths are mostly consistent with the geometrically predicted paths as shown in Fig.~\ref{subfig:SchematicDepictionHallWay2D} using a known environment geometry. Fig.~\ref{fig:realDataAgent} presents the agent position estimation error versus time $n$. The results indicate that \ac{prop} accurately localizes the agent, achieving an average agent position estimation error of $ 5.4 $\,cm, despite \ac{olos} conditions after step $n=22$.

%% file: InputFiles/Conclusions.tex
In this paper, we introduced a multipath-based \ac{slam} method for \ac{dmimo} systems that performs adaptive data fusion across \acp{mpc} and \acp{bs}. The main contribution is a novel statistical model that integrates amplitude statistics of propagation paths with an \ac{mva} model for representing reflective surfaces, enabling automatic adaptation to time-varying detection probabilities and measurement variances. The detection of \acp{mva} and propagation paths is jointly formulated within a unified Bayesian inference framework, enabling ``soft'' ray tracing and the estimation of the number and states of \acp{mva} and propagation paths. Furthermore, we develop an improved initialization procedure for emerging surfaces, which enables the reliable introduction of new map features even when measurements originate solely from double-bounce paths. These aspects result in a highly flexible and robust data fusion approach for multipath-based \ac{slam} in dynamic and complex environments while maintaining moderate computational complexity. Numerical results based on synthetic and real data demonstrate that the proposed method significantly improves localization and mapping performance compared to state-of-the-art multipath-based \ac{slam} methods based on \ac{va} features, particularly under \ac{olos} conditions. Our results also indicate that the proposed method can approach the \acp{pcrlb}. They furthermore highlight the potential of the proposed framework beyond mapping and localization, for example, in predicting channel state information for integrated sensing and communication systems.

Promising directions for future research include extending the current work to coherent data fusion across \acp{pa} while accounting for system impairments, e.g., imperfect time and phase synchronization \cite{FasDeuKesWilColWitLeiSecWym:STSP2025}; employing a sigma point-based implementation \cite{KimGranSveKimWym:TVT2022, AnnaFusion2025} and parallel computation on GPUs to achieve real-time performance; extending to multiple measurement-to-feature \ac{da} and feature models to account for measurements originating from rough surfaces or other complex structures \cite{WieVenWilLei:JAIF2023,WieVenWilWitLei:Fusion2024,ZhaFanGaoDai:TWC2025}, and using particle flow to improve inference in highly nonlinear system models with high-dimensional states \cite{JanMeySnyWigBauHil:J23,ZhaMey:TSP2024,WieLeiMeyWit:TSIPN2023}; extending to a hybrid inference framework, e.g., neural-enhanced belief propagation \cite{LiaMey:TSP2024, WeiLiaMey:FUSION2024, VenLeiTerWit:JSP2023}, which combines model-based and data-driven approaches, allowing statistical models to leverage information learned from raw sensor data; and extending to differentiable ray tracing \cite{JakobDiffRT2024}, enabling the learning of differentiable parameters associated with the \ac{rf} environment and system in a physically interpretable manner.